%% file: lensing-RAR.tex
\documentclass[a4paper,11pt]{article}
\pdfoutput=1 %

\usepackage{jcappub} %

\usepackage[T1]{fontenc} %

\title{Radial acceleration relation of galaxies with joint kinematic and weak-lensing data}

\author[a,1]{T. Mistele,\note{Corresponding author.}}
\author[a]{S. McGaugh,}
\author[b]{F. Lelli,}
\author[c]{J. Schombert}
\author[d]{and P. Li}

\affiliation[a]{Department of Astronomy, Case Western Reserve University,\\10900 Euclid Avenue, Cleveland, Ohio 44106, USA}
\affiliation[b]{INAF – Arcetri Astrophysical Observatory,\\Largo Enrico Fermi 5, 50125 Firenze, Italy}
\affiliation[c]{Department of Physics, University of Oregon,\\Eugene, Oregon 97403, USA}
\affiliation[d]{Leibniz-Institute for Astrophysics,\\ An der Sternwarte 16, 14482 Potsdam, Germany}

\emailAdd{tobias.mistele@case.edu}
\emailAdd{stacy.mcgaugh@case.edu}
\emailAdd{federico.lelli@inaf.it}
\emailAdd{jschombe@uoregon.edu}
\emailAdd{pli@aip.de}

\abstract{
  We combine kinematic and gravitational lensing data to construct the Radial Acceleration Relation (RAR) of galaxies over a large dynamic range.
  We improve on previous weak-lensing studies in two ways.
  First, we compute stellar masses using the same stellar population model as for the kinematic data.
  Second, we introduce a new method for converting excess surface density profiles to radial accelerations.
  This method is based on a new deprojection formula which is exact, computationally efficient, and gives smaller systematic uncertainties than previous methods.
  We find that the RAR inferred from weak-lensing data smoothly continues that inferred from kinematic data by about $2.5\,\mathrm{dex}$ in acceleration.
  Contrary to previous studies, we find that early- and late-type galaxies lie on the same joint RAR when a sufficiently strict isolation criterion is adopted and their stellar and gas masses are estimated consistently with the kinematic RAR.
}

\ExplSyntaxOn
\cs_new:Npn \expandableinput #1
{ \use:c { @@input } { \file_full_name:n {#1} } }
\AddToHook{env/tabular/begin}
{ \cs_set_eq:NN \input \expandableinput }
\ExplSyntaxOff

\newcommand{\gobs}{\ensuremath{g_{\mathrm{obs}}}}
\newcommand{\gbar}{\ensuremath{g_{\mathrm{bar}}}}

\begin{document}
\maketitle
\flushbottom

\section{Introduction}
\label{sec:introduction}

A well-known property of spiral galaxies is that their rotation curves are approximately flat at large radii \citep{Rubin1978,Bosma1981}.
This flatness is, however, far from the only regularity contained in the dynamical properties of spiral galaxies.
Indeed, a lot of additional regularities are captured by scaling relations like the Baryonic Tully-Fisher Relation \citep[BTFR,][]{McGaugh2000,McGaugh2016,Lelli2016c,Lelli2019,Schombert2020}, the Central Density Relation \citep[CDR,][]{Lelli2016b,Milgrom2016b}, and the Radial Acceleration Relation \citep[RAR,][]{Sanders1990,McGaugh2004b,Wu2015,McGaugh2016b,Lelli2017b,Shelest2020}.
The BTFR links the baryonic mass $M_b$ (stars and gas) of a galaxy to its flat rotation speed\footnote{The BTFR is a generalization of the original Tully-Fisher relation \citep{Tully1977} between luminosity and line-width.
These quantities are proxies for stellar mass and rotation speed \citep{Verheijen2001,Ponomareva2017,Ponomareva2018,Lelli2019}.}.
The CDR links the baryonic surface density at the galaxy center ($R\rightarrow 0$) with the dynamical surface density inferred from the inner steepness of the rotation curve.
The RAR links the observed centripetal acceleration at each radius ($\gobs = V_c^2/R$) with that predicted from the observed distribution of stars and gas assuming Newtonian gravity ($\gbar = |\partial \Phi_{\mathrm{bar}}/\partial R|$).
In particular, one finds $\gobs \approx \gbar$ at large accelerations, $\gbar \gg a_0$, and $\gobs \approx \sqrt{a_0 \gbar}$ at small accelerations, $\gbar \ll a_0$.
The acceleration scale $a_0$ occurs in all three scaling relations (BTFR, CDR, RAR).
It has a universal value of about $10^{-10}\,\mathrm{m/s}^2$ despite playing a different physical role in each relation.

Scaling relations like the RAR are commonly measured using radio interferometry of the $21\,\mathrm{cm}$ spin-flip transition of atomic hydrogen. Such observations probe over many tens of $\mathrm{kpc}$ \citep{Lelli2016}, sometimes reaching $100\,\mathrm{kpc}$ \citep{Noordermeer2005,Lelli2010}, without revealing any credible deviation from a universal RAR.
Another probe is offered by weak gravitational lensing \citep[e.g.,][]{Brimioulle2013,Milgrom2013,Brouwer2017}.
Recently, it was shown in Ref.~\cite{Brouwer2021} that lensing observations probe to at least $300\,\mathrm{kpc}$ and perhaps to $\mathrm{Mpc}$ scales.

Here, we rederive the RAR from the KiDS DR4 weak-lensing data \citep{Kuijken2019,Giblin2021,Bilicki2021} that was also used in Ref.~\cite{Brouwer2021},
  improving key aspects of the analysis.
In particular, we estimate stellar masses using the same stellar population synthesis model \citep{Schombert2014} that was used for previous determinations of the RAR \citep{Lelli2016,Lelli2017b}.
We further introduce a new method for converting weak-lensing data to radial accelerations.
This method is based on a new, exact deprojection formula that relates excess surface densities and radial accelerations.

We first introduce our new deprojection formula and the corresponding method for obtaining radial accelerations from weak-lensing data in Sec.~\ref{sec:newmethod}.
In Sec.~\ref{sec:data} and Sec.~\ref{sec:masses}, we then describe, respectively, the data set and baryonic mass estimates we use.
We investigate the resulting radial acceleration in Sec.~\ref{sec:results}.
After a brief discussion in Sec.~\ref{sec:discussion} we conclude in Sec.~\ref{sec:conclusion}.

\section{From excess surface density to acceleration}
\label{sec:newmethod}
\label{sec:newmethod:individual}

Weak-lensing observations measure the distortion of source images due to massive objects, called lenses, along the line-of-sight.
Here, we consider galaxy-galaxy lensing where both the sources and lenses are galaxies.
The goal is to measure the radial accelerations $g_{\mathrm{obs}}$ around the lens galaxies.
The main ingredients for this approach are the ellipticities of the source galaxies.
These contain contributions from both intrinsic ellipticities and distortions induced by gravitational lensing.
To isolate the effect of gravitational lensing, one averages over many source-lens pairs, after which the intrinsic ellipticities average out.
In this way, one obtains an averaged tangential shear profile $\gamma_t(R)$ as a function of the projected distance from the lens $R$, which is related to the so-called excess-surface density (ESD) $\Delta \Sigma(R)$ \citep{Kaiser1995,Bartelmann1995}.
From such ESD profiles, one can then obtain averaged gravitational accelerations.

In this section, we introduce a new method for obtaining the radial acceleration $g_{\mathrm{obs}}$ of an individual galaxy given its ESD profile $\Delta \Sigma$.
Following Ref.~\cite{Brouwer2021}, we assume General Relativity (GR) and the usual thin-lens approximation \citep{Bartelmann2001}, allowing for both baryonic and dark matter.
Nevertheless, our results have a straightforward interpretation also in some modified gravity models;
namely in models where lensing works as in GR, just with the lenses having a specific total mass $M(r)$ (see below).
Examples are discussed in Refs.~\cite{Mistele2023,Mistele2023c,Milgrom2013}.

The ESD profile $\Delta \Sigma(R)$ of an individual galaxy is defined in terms of its surface density $\Sigma(R)$,
\begin{align}
 \label{eq:esd}
 \Delta \Sigma(R) \equiv \frac{2}{R^2} \int_0^R dR' R' \Sigma(R') - \Sigma(R) \,.
\end{align}
Galactic scaling relations like the RAR are based on accelerations.
Thus, to test these relations using weak-lensing data, we need to convert ESD profiles to accelerations.
Here, we assume spherical symmetry which is a reasonable approximation at the large radii involved (up to a few $\mathrm{Mpc}$, see below).
The relevant acceleration is then the radial acceleration $g_{\mathrm{obs}}(r) = G M(r)/r^2$, where $M(r)$ is the cumulative 3D mass calculated from the total density profile $\rho(r)$, including both baryons and dark matter, and $r$ is the spherical radius.

In Ref.~\cite{Brouwer2021}, two methods for converting ESD profiles to accelerations were proposed.
The first is called the ``SIS'' method which assumes that the total density profile $\rho$ of the galaxy is a singular isothermal sphere (SIS).
With this assumption, one obtains the simple relation
\begin{align}
 \label{eq:gobs_sis}
 \left. g_{\mathrm{obs}}(R) \right|_{\mathrm{SIS}} = 4 G \Delta \Sigma(R) \,.
\end{align}
The second method is called the ``PPL'' method and is more elaborate.
One fits a piecewise power-law density profile $\rho$ to the observed ESD profile and, from the density profile obtained in this way, one finds the acceleration by integration.
The advantage of the PPL method is that it does not assume a specific density profile.
It does, however, come with a much higher computational cost and is more complicated to implement.

Here, we introduce a new method for converting ESD profiles to accelerations.
We use a new, exact deprojection formula which relates the ESD profile $\Delta \Sigma$ and the radial acceleration $g_{\mathrm{obs}}$.
We make no assumptions about the density profile $\rho$ except that it is spherically symmetric and asymptotically falls off faster than $1/r$ with the spherical radius\footnote{
  In particular, we do not assume that $\rho$ is positive.
  Thus, our results apply also to negative (effective) densities.
  This is relevant, for example, for Aether Scalar Tensor Theory \citep{Skordis2020} where the ghost condensate density may be negative \citep{Mistele2023} or for modified gravity theories with negative ``phantom dark matter'' densities \citep{Milgrom1986b}.
} $r$.
Despite this generality, the method is simple to implement and computationally efficient.
The basic relation is derived in Appendix~\ref{sec:appendix:derivation} and reads
\begin{align}
 \label{eq:gobs_from_esd}
 \left.g_{\mathrm{obs}}(R)\right|_{\mathrm{exact}} = 4G \int_0^{\pi/2} d\theta \, \Delta \Sigma\left(\frac{R}{\sin \theta}\right) \,.
\end{align}
It is instructive to compare this new relation to the ``SIS'' relation Eq.~\eqref{eq:gobs_sis}.
We see that the SIS method obtains the radial acceleration $g_{\mathrm{obs}}$ at a radius $R$ from the ESD profile at the same radius $R$.
Similarly, Eq.~\eqref{eq:gobs_from_esd} places most weight on radii near $R$, but also averages over radii larger than $R$.

In particular, the integral in Eq.~\eqref{eq:gobs_from_esd} involves $\Delta \Sigma(R)$ at arbitrarily large radii $R$.
This is entirely expected:
The ESD $\Delta \Sigma$ is defined in terms of the surface density $\Sigma$ which is an integral over the density $\rho$ that extends to infinity.
In contrast, $g_{\mathrm{obs}}(r)$ does not depend on $\rho$ at arbitrarily large radii.
Thus, getting $g_{\mathrm{obs}}$ from $\Delta \Sigma(R)$ requires information about $\Delta \Sigma(R)$ out to infinity.

In practice, observations of the ESD profile $\Delta \Sigma$ extend only to a certain maximum radius $R_{\mathrm{max}}$.
So we need to make a choice about what to do at radii larger than $R_{\mathrm{max}}$.
Here, we assume that, at radii larger than $R_{\mathrm{max}}$, the ESD profile $\Delta \Sigma$ behaves as a singular isothermal sphere, i.e. $\Delta \Sigma \propto 1/R$,
\begin{align}
 \Delta \Sigma(R > R_{\mathrm{max}}) \equiv \Delta \Sigma(R_{\mathrm{max}}) \frac{R_{\mathrm{max}}}{R} \,.
\end{align}
Other choices are possible.
But without appealing to a specific theoretical model of how the density $\rho$ should typically look like in our universe, we have no good reason to choose one over the other.
Thus, there is a systematic uncertainty corresponding to how $\Delta \Sigma$ behaves\footnote{
  Implicitly, both the ``SIS'' and ``PPL'' methods also depend on such choices.
  The SIS method assumes a singular isothermal sphere.
  The PPL method assumes that the slope of the density $\rho$ is constant between the last data point and infinity.
} at radii larger than $R_{\mathrm{max}}$.
We quantify it by calculating $g_{\mathrm{obs}}$ using two opposite, extreme assumptions and then taking the difference between these two versions of $g_{\mathrm{obs}}$.
Here, we consider the two assumptions that $\Delta \Sigma$ drops to zero at the last data point and that $\Delta \Sigma$ keeps the value it has at $R_{\mathrm{max}}$ up to infinity,
\begin{align}
 \Delta \Sigma(R > R_{\mathrm{max}}) = 0 \,, \quad
 \Delta \Sigma(R > R_{\mathrm{max}}) = \Delta \Sigma(R_{\mathrm{max}}) \,.
\end{align}
That is, written symbolically, we adopt the following systematic error due to this uncertainty at large radii,
\begin{align}
 \label{eq:syst_err_gobs_extrapolate}
 \left.\sigma_{g_{\mathrm{obs}}}(R)\right|^{\mathrm{extrapolate}}_{\mathrm{systematic}} = \left| g_{\mathrm{obs}}(R) |_{\mathrm{extrapolate\;flat}} - g_{\mathrm{obs}}(R) |_{\mathrm{extrapolate\;0}} \right| \,.
\end{align}
In practice, this systematic error is often negligible.
It only becomes important close to $R_{\mathrm{max}}$.
This is for two reasons.
First, the integrand $\Delta \Sigma$ in the integral Eq.~\eqref{eq:gobs_from_esd} drops off with radius.
Second, irrespective of the integrand, the integral Eq.~\eqref{eq:gobs_from_esd} gives more weight to radii close to $R$ and only relatively little weight to arbitrarily large radii.

Another systematic uncertainty arises because, in practice, we know the ESD profile $\Delta \Sigma(R)$ only in discrete radial bins (Appendix~\ref{sec:appendix:newmethod:observational}), while the integral in the deprojection formula Eq.~\eqref{eq:gobs_from_esd} is continuous.
So we have to interpolate between the discrete radial bins.
This also applies to the integral Eq.~\eqref{eq:stat_err_gobs} for the statistical errors we introduce below.
Here, for simplicity, we linearly interpolate between the discrete radial bins.
To estimate the systematic uncertainty associated with this interpolation, we calculate $\gobs$ with both linear and quadratic interpolation and take the difference as an additional systematic error.
Symbolically,
\begin{align}
 \label{eq:syst_err_gobs_interpolate}
 \left.\sigma_{g_{\mathrm{obs}}}(R)\right|_{\mathrm{systematic}}^{\mathrm{interpolation}} = \left| g_{\mathrm{obs}}(R)|_{\mathrm{linear}} - g_{\mathrm{obs}}(R)|_{\mathrm{quadratic}} \right| \,.
\end{align}

Below, we will see that these systematic uncertainties are typically both small compared to the statistical uncertainties (except close to the last data point at $R_{\mathrm{max}}$).
This should be contrasted with the systematic uncertainties of the SIS and PPL methods.
For them, a systematic uncertainty of $0.05\,\mathrm{dex}$ on $g_{\mathrm{obs}}$ was estimated in Ref.~\cite{Brouwer2021}.\footnote{
  In the text of Ref.~\cite{Brouwer2021}, this is described as a $0.1\,\mathrm{dex}$ systematic error.
  But from Fig.~2 of Ref.~\cite{Brouwer2021} it can be seen that $0.05\,\mathrm{dex}$ is a better estimate (Brouwer, priv. comm.).
  Using our exact result, we have verified that $0.05\,\mathrm{dex}$ is indeed a reasonable estimate.
}
This is comparable to the statistical uncertainties found in Ref.~\cite{Brouwer2021}.

For the statistical errors, we assume that the measurements of $\Delta \Sigma$ at different radii are independent.
We further discuss this assumption in Appendix~\ref{sec:appendix:errors}.
The statistical errors on $g_{\mathrm{obs}}$ are then given by (see Eq.~\eqref{eq:gobs_from_esd}),
\begin{align}
 \label{eq:stat_err_gobs}
  \left.\sigma^2_{g_{\mathrm{obs}}}(R)\right|_{\mathrm{statistical}} = (4G)^2 \int_0^{\pi/2} d\theta \, \sigma^2_{\Delta \Sigma}\left(\frac{R}{\sin \theta}\right) \,,
\end{align}
where $\sigma_{\Delta \Sigma}$ denotes the statistical error on $\Delta \Sigma$.
This formula requires the statistical uncertainties of the ESD profile $\Delta \Sigma$ at radii beyond the last data point.
However, these uncertainties are systematic ones that we treat separately.
Thus, beyond the last radial bin $R_{\mathrm{max}}$, we adopt
\begin{align}
\sigma_{\Delta \Sigma}(R > R_{\mathrm{max}}) \equiv 0 \,.
\end{align}
More specifically, we set $\sigma_{\Delta \Sigma}$ to zero beyond the last bin edge, not beyond the last bin center.
Otherwise, the $\theta$ integral would give exactly zero already at the last bin center.
Between the last bin center and the last bin edge we interpolate linearly.

So far we have considered ESD profiles and radial accelerations as a function of radius $R$.
Below, we will mainly be interested in the RAR that requires $g_{\mathrm{obs}}$ as a function of the Newtonian baryonic acceleration $\gbar \equiv G M_b(R)/R^2$.
For an individual galaxy, obtaining $\gobs(\gbar)$ from $\gobs(R)$ is straightforward, at least as long there is a one-to-one mapping between radii $R$ and baryonic accelerations $\gbar$.
Indeed, instead of $\gobs(R)$ one can simply use $\gobs(R(\gbar))$ where the function $R(\gbar)$ maps values of $\gbar$ to values of $R$ for that galaxy.

Below, we will not be interested in the lensing signal of an individual galaxy but in the stacked signal of a large number of galaxies.
In Appendix~\ref{sec:appendix:newmethod}, we explain how to adapt the method discussed above to this practically relevant case of stacking a large number of galaxies.
We also explain in Appendix~\ref{sec:appendix:newmethod} how we obtain the required ESD profiles from observational data.

There are some subtleties when adapting our method to work with stacked data.
Basically, there is a choice between first deprojecting individual galaxies using Eq.~\eqref{eq:gobs_from_esd} and then stacking (which is slightly more susceptible to observational systematics) versus first stacking individual galaxies and then deprojecting (in which case the deprojection is not exact).
In practice, both choices give almost identical results (see Appendix~\ref{sec:appendix:randomESD}) despite their different systematics.
Thus, we trust our results despite these subtleties.
The results shown below are obtained using the first method, i.e. we first deproject and then stack.
We discuss this in detail in Appendix~\ref{sec:appendix:newmethod}.

We note that the deprojection formula Eq.~\eqref{eq:gobs_from_esd} can also be used to infer cumulative 3D mass profiles $M(<R)$ from weak-lensing observations.
Indeed, in spherical symmetry we have $M(< R) = (R^2/G) \, g_{\mathrm{obs}}(R)$.
Thus, Eq.~\eqref{eq:gobs_from_esd} may also be useful when one is not directly interested in radial accelerations.

\section{Data}
\label{sec:data}

We adopt $H_0 = 73\,\mathrm{km}\,\mathrm{s}^{-1}\,\mathrm{Mpc}^{-1}$ for consistency with the RAR derived from kinematic measurements \citep{Lelli2017b} and the Hubble constant measured with the BTFR \citep{Schombert2020}.
Below we sometimes use the notation $h_{70}$, defined as $H_0/(70\,\mathrm{km}\,\mathrm{s}^{-1}\,\mathrm{Mpc}^{-1})$.
Our choice corresponds to $h_{70} = 73/70$.
We adjust our masses and radii accordingly.

We use mostly the same data as Ref.~\cite{Brouwer2021}.
Specifically, we use source galaxies from the KiDS-1000 SOM-gold catalog \citep{Kuijken2019,Wright2020,Giblin2021,Hildebrandt2021} and lens galaxies from the KiDS-bright sample \citep{Bilicki2021}.
We use the best-fit photometric redshifts $z_{B,s}$ for the sources $s$, and the redshifts $z_{\mathrm{ANN},l}$ from the machine-learning method ``\textsc{ANNz}2'' \citep{Sadeh2016} for the lenses $l$.
We include only lenses with `masked = 0` and with redshift $0.1 < z_{\mathrm{ANN},l} < 0.5$.

In addition, we restrict the lens galaxies to be isolated using the criterion introduced by Ref.~\cite{Brouwer2021}.
That is, for each lens, we enforce a lower bound $R_{\mathrm{isol}}$ on the 3D distance to the closest neighboring galaxy with at least $10\%$ of its stellar mass.
We use $R_{\mathrm{isol}} = 4\,\mathrm{Mpc}/h_{70}$ unless stated otherwise.
This is larger than the value $R_{\mathrm{isol}} = 3\,\mathrm{Mpc}/h_{70}$ used by Ref.~\cite{Brouwer2021}.
We explore the effects of making this criterion stricter or laxer in Sec.~\ref{sec:etgltg} below.
We use the stellar masses of Ref.~\cite{Brouwer2021} for the isolation criterion.
We do this for simplicity and to have a direct comparison to Ref.~\cite{Brouwer2021}.
Adopting our own mass estimates from Sec.~\ref{sec:masses} instead would not affect our conclusions, as we explain there.

Note that the isolation criterion is of great practical importance.
Non-isolated galaxies pick up signals from surrounding structure which violates the assumption of spherical symmetry that we make; see also the discussion of the so-called two-halo term in Ref.~\cite{Brouwer2021}.
Even if we ignore this assumption of spherical symmetry, our analysis still requires isolated galaxies because we are not interested in the structure surrounding lens galaxies but in the intrinsic properties of the lens galaxies themselves.

Since the isolation criterion operates on 3D distances it depends on the redshifts of the KiDS-bright lenses.
There are no spectroscopic redshifts available for this sample, so we use the photometry-based redshifts $z_{\mathrm{ANN},l}$ which, as all photometric redshifts, have non-negligible uncertainties.
We discuss this more in Sec.~\ref{sec:discussion} below. 
An alternative would be to base the isolation criterion on projected distances $R_{\mathrm{isol,proj}}$ instead of 3D distances $R_{\mathrm{isol}}$.
This would give a stricter criterion which does not rely on photometric redshifts.
We nevertheless prefer the isolation criterion based on 3D distances for two reasons.
First, this allows for a direct comparison to Ref.~\cite{Brouwer2021}.
Second, the sample size becomes very small even for moderate values of $R_{\mathrm{isol,proj}}$.
For example, for $R_{\mathrm{isol,proj}} = 1\,\mathrm{Mpc}/h_{70}$, there are only $196$ lenses left.

Following Ref.~\cite{Brouwer2021}, we impose an upper limit on the stellar masses of the lens galaxies because massive galaxies have significantly more satellites \citep[see also Sect. 2.2.3 of Ref.][]{Brouwer2017} and because our adopted light to stellar-mass conversions \cite{Schombert2014}
(see Sec.~\ref{sec:masses}) may 
require modest revision for the most massive ETGs \cite{Schombert2019}
(we discuss this more in Sec.~\ref{sec:massbins}).
We adopt $\log_{10} M_\ast/M_\odot < 11.1$.
For this, we use our stellar mass estimates from Sec.~\ref{sec:masses}.
This leaves 106,843 lenses in our sample which is less than the 259,383 used by Ref.~\cite{Brouwer2021}.
The biggest difference comes from the stricter isolation criterion we apply.
Since our stellar masses are, on average, larger than those of Ref.~\cite{Brouwer2021}, see Sec.~\ref{sec:masses}, we use a slightly larger cut-off for $\log_{10} M_\ast/M_\odot$, namely of $11.1$ instead of $11.0$.

For the sources, we restrict the SOM-gold catalog to `SG\_FLAG = 1', `SG2DPHOT = 0', and `CLASS\_STAR $<$ 0.5' in order to remove stars from the sample \citep{Brouwer2021,Kuijken2019}.
We also apply the quality cut `IMAFLAGS\_ISO = 0' and the recommended mask `MASK \& 28668 = 0' \citep{Brouwer2021,Kuijken2019}.
In addition, when estimating the tangential shear around a lens, we follow Ref.~\cite{Brouwer2021} and consider only sources $s$ that are not too close to a given lens $l$ in redshift to avoid including sources in front of the lens $l$.
Specifically, we consider only sources with $z_{B,s} > z_{\mathrm{ANN},l} + 0.2$.

For the critical surface density (see Appendix~\ref{sec:appendix:newmethod:observational} for where this is needed), we use the method of Refs.~\cite{Dvornik2017,Brouwer2021} to take into account uncertainties in  the redshifts.
That is, we adopt
\begin{align}
 \label{eq:sigmacrit_full}
 \Sigma^{-1}_{\mathrm{crit},ls} = \frac{4 \pi G}{c^2} \int_0^{\infty} dz_l \, p_{z_{\mathrm{ANN},l}}(z_l) \cdot D(z_l) \int_{z_l}^{\infty} d z_s \, n_{z_l,z_{B,s}}(z_s) \cdot \frac{D(z_l, z_s)}{D(z_s)}  \,.
\end{align}
Here, $D(z_l)$ and $D(z_s)$ are the angular diameter distances to the lens and source, respectively, and $D(z_l, z_s)$ is the angular diameter distance between the lens and the source.
Following Ref.~\cite{Brouwer2021}, we calculate these distances assuming a flat FLRW cosmology with $\Omega_{\mathrm{m}} = 0.2793$.
Further, $p_{z_{\mathrm{ANN},l}}(z)$ is a normal distribution centered on the lens redshift $z_{\mathrm{ANN},l}$ with standard deviation $\sigma = 0.02\cdot(1+z_{\mathrm{ANN},l})$.
The function $n_{z_l,z_{B,s}}(z)$ is determined as follows.
For a given value of the integration variable $z_l$ and a given best-fit photometric source redshift $z_{B,s}$, we first find out in which of the five tomographic bins from Ref.~\cite{Hildebrandt2021} the $z_{B,s}$ value belongs and get the corresponding redshift distribution function from Ref.~\cite{Hildebrandt2021}.
Then we normalize this distribution to unity in the interval $[z_l, \infty)$.

\section{Masses and mass distributions}
\label{sec:masses}

\begin{figure}
 \centering
  \includegraphics[width=.8\textwidth]{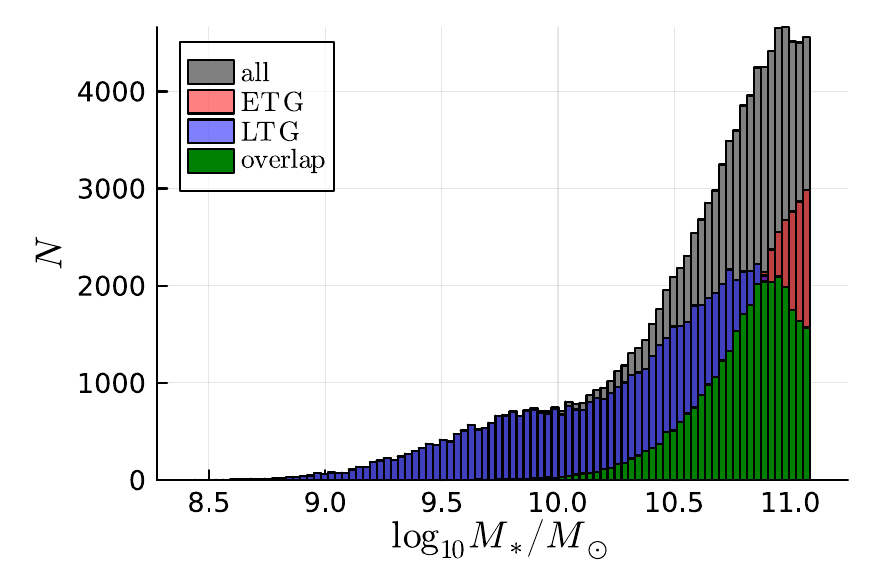}
 \caption{
   The stellar mass histogram assuming our stellar masses for the whole sample (gray), for early type galaxies (ETGs in red), for late type galaxies (LTGs in blue), and for the overlap of ETGs and LTGs (green).
 }
 \label{fig:Mstar}
\end{figure}

In order to properly compare kinematic and lensing data, it is important that the stellar mass informing \gbar\ 
is consistent for both samples. We do not observe stellar mass directly, but rather the apparent magnitude
of each galaxy. This is converted to a luminosity for the assumed distance scale, and then to a stellar mass using a mass-to-light ratio indicated by a stellar population model informed by the observed colors. Stellar population models have converged so that most give similar results, but agreement is never perfect and systematic differences still occur at the factor of two level \citep{McGaugh2014}.

To insure a uniform stellar mass scale, we have reanalyzed the KiDS data used here with the same stellar population synthesis model \citep{Schombert2014} used for the kinematic data \citep{Lelli2017b}. Comparing our results to those of Ref.~\cite{Brouwer2021}, we find good agreement overall, albeit with a slight morphology dependence. Following Ref.~\cite{Brouwer2021}, we define early and late type galaxies (ETGs and LTGs) based on the color\footnote{Specifically, we use the apparent magnitudes `MAG\_GAAP\_u' and `MAG\_GAAP\_r' provided by KiDS DR4 \citep{Kuijken2019}.}  split $u-r \gtrless 2.5$. Indeed, the agreement is uniform enough that the bulk of the sample can be described as a simple multiplicative scaling factor, which we parameterize by defining the ratio $Q$ such that\footnote{
  The values for $\log_{10} M_\ast^{\mathrm{KiDS}}/M_\odot$ used by Ref.~\cite{Brouwer2021} can be obtained from the the KiDS bright data tables as `MASS\_BEST + (MAG\_GAAP\_r - MAG\_AUTO\_CALIB)/2.5 + 0.056'.
  This assumes $h_{70} = 1$.
  We correct these values to account for our choice, $h_{70} = 73/70$.
} $M_\ast = Q M_\ast^{\mathrm{KiDS}}$.
The agreement in the average mass-to-light ratio is excellent for LTGs: $Q_{\mathrm{LTG}}=1.0$.
For ETGs, we find a modest offset: $Q_{\mathrm{ETG}} = 1.4$. This goes in the sense that we assess ETGs to be slightly more massive than do Ref.~\cite{Brouwer2021}.
We apply these average correction factors to each individual galaxy depending on which type they are.
We show the resulting stellar masses in Fig.~\ref{fig:Mstar}.

An offset like the one we find for ETGs is to be expected when comparing different stellar population synthesis models. Indeed, the exact same type-dependent factor is found in Ref.~\cite{Brouwer2021} when comparing stellar masses for ETGs in their KiDS and GAMA\footnote{We have previously found the stellar population models applied here to be in good agreement with those of GAMA \citep{Pengfei2019}.} samples.
The building block of any composite stellar population model is a single stellar population (SSP) of a given age: we use the SSPs from Ref.~\cite{CG2010}, while the KiDS team used the older SSPs from Ref.~\cite{BC2003}. In addition, the composite stellar population models we employ \citep{Schombert2014} assume physically-motivated differences in the star-formation and chemical-enrichment histories of early and late types: ETGs form most of their stars early and quickly reach high metallicities, whereas LTGs have a more continuous star formation and a more gradual chemical enrichment \citep[see also][]{Schombert2019, Schombert2022}. Unfortunately, Ref.~\cite{Brouwer2021} does not provide details about the star-formation and chemical-enrichment histories of their stellar population models, so we cannot discern what is the dominant cause of disparity in the stellar masses of ETGs.
Two possible culprits are the treatment of AGB stars, for which we use an empirical calibration \cite{Schombert2014}, and metallicity, for which we take care to use a distribution of stellar metallicities that is consistent with chemical evolution models rather than the usual approximation that all stars have the same metallicity. The metallicity plays an important role in the color-magnitude relation of ETGs \cite{Kodama1997} that may induce larger mass-to-light ratios in brighter ETGs \cite{Schombert2019}. 
Though the difference is modest ($\sim$40$\%$), it is important to whether or not ETGs follow the same RAR as LTGs.

We follow Ref.~\cite{Brouwer2021} in using scaling relations to account for the baryonic mass in gas.
Massive ETGs, such as those considered here (Fig.~\ref{fig:Mstar}), are known to have X-ray-emitting coronae of hot ionized gas \citep[e.g.,][]{Buote2012}.
Thus, we add hot gas to the total baryonic mass according to the scaling relation \citep{Chae2021b},
\begin{align}
 \label{eq:fhot}
 \frac{M_{g,\mathrm{hot}}}{M_\ast} = 10^{-5.414} \cdot \left(\frac{M_\ast}{M_\odot}\right)^{0.47} \,.
\end{align}
Star-forming LTGs have a non-negligible interstellar medium of atomic and molecular gas, so we add cold gas according to the scaling relation
\begin{align}
 \label{eq:fcold}
 \frac{M_{g,\mathrm{cold}}}{M_\ast} = \frac{1}{X} \left(11550 \left(\frac{M_\ast}{M_\odot}\right)^{-0.46} + 0.07 \right) \,,
\end{align}
where
\begin{align}
 \label{eq:metal}
 X = 0.75  - 38.2 \left(\frac{M_\ast}{1.5\cdot10^{24}M_\odot}\right)^{0.22} \,.
\end{align}
The first term in Eq.~\eqref{eq:fcold} represents atomic gas according to the scaling relation from Ref.~\cite{Lelli2016} with $M_\ast = 0.5 L_{[3.6]}$.
The second term takes into account molecular gas.
Eq.~\eqref{eq:metal} accounts for the variation of the hydrogen fraction $X$ as metallicity varies with stellar mass \citep{McGaugh2020b}.

This accounting of the cold gas is practically indistinguishable from the relation adopted by Ref.~\cite{Brouwer2021} over the relevant mass range.
Thus, our baryonic mass estimate for LTGs is close to that of Ref.~\cite{Brouwer2021}.
For ETGs, we use the hot gas mass estimate Eq.~\eqref{eq:fhot}, while in Ref.~\cite{Brouwer2021} the same cold gas mass estimate as for LTGs is used.
At large stellar masses, our hot gas mass estimate is larger than the cold gas mass estimate of Ref.~\cite{Brouwer2021}.
Thus, since ETGs tend to have large stellar masses (Fig.~\ref{fig:Mstar}), our baryonic mass estimate for most ETGs is larger than that of Ref.~\cite{Brouwer2021}.
On average,  our ETG baryonic masses are larger by a factor of about $1.7$, which includes the effects of our different stellar $M_\ast/L$, our different gas mass estimate, and our different choice of $H_0$.
Our results below mainly depend on the total baryonic mass, not on how this mass is distributed between gas and stars.
Thus, we would expect to find similar results with $Q_{\mathrm{ETG}} = 1.7$ and the gas mass estimate of Ref.~\cite{Brouwer2021}.

For simplicity, we treat both the stellar mass and the gas mass as point masses.
This assumption is sensible because the bulk of the observed baryonic mass is typically contained within tens of $\mathrm{kpc}$, which is comparable to the smallest radial bins we consider while the largest exceed $3\,\mathrm{Mpc}$.
Atomic gas disks in LTGs and hot gas halos in ETGs are sometimes observed to extend out to $\sim 100\,\mathrm{kpc}$, but both components typically give a minor gravitational contribution out to these radii.
Hot gas around galaxies may be more extended than that, forming  the so-called circumgalactic medium, but its amount and distribution are poorly constrained, so they will be neglected in our work.
In Appendix~\ref{sec:appendix:SIShotgas}, we explicitly show that modeling the hot gas from Eq.~\eqref{eq:fhot} as an extended distribution instead of a point mass does not significantly change our results.

As discussed in Sec.~\ref{sec:data}, we reuse the stellar masses of Ref.~\cite{Brouwer2021} for the isolation criterion defined there.
We expect that using our own stellar masses instead would not significantly affect our results, because the isolation criterion considers only ratios of stellar masses which are mostly unchanged by the simple rescaling we adopt here.
The only reason our rescaled stellar masses would make a difference at all is that we rescale masses of ETGs and LTGs differently.
Thus, when an LTG lens has an ETG neighbor or vice versa, the isolation criterion would change when adopting our stellar masses.
In practice, however, this does not change any of our conclusions for the following reason.
Our stellar mass estimates agree with those from Ref.~\cite{Brouwer2021} for LTGs and are a bit higher than those from Ref.~\cite{Brouwer2021} for ETGs.
Thus, for ETGs we are actually adopting a stricter isolation criterion by sticking with the stellar masses from Ref.~\cite{Brouwer2021}.
For LTGs, our isolation criterion would be stricter if we used our own stellar masses instead.
But in Sec.~\ref{sec:etgltg}, we will see that our LTG sample is sufficiently isolated anyway, so this is not a problem.
We also note that the precise quantitative meaning of $R_{\mathrm{isol}}$ is not important to any of our results.
It is only important that by adjusting $R_{\mathrm{isol}}$ we can make the isolation criterion more or less strict.

\section{Results}
\label{sec:results}

\subsection{The radial acceleration relation}
\label{sec:rar}

\begin{figure}
 \centering
  \includegraphics[width=.9\textwidth]{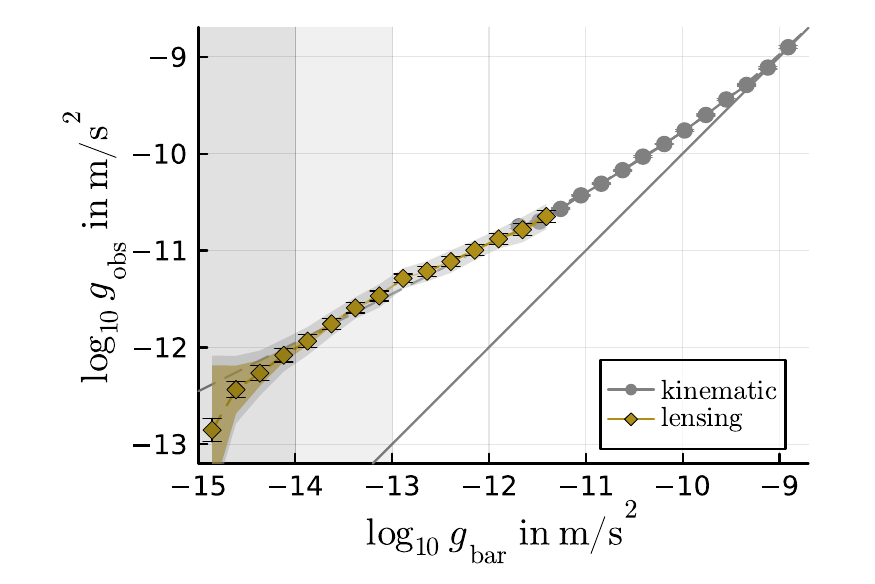}
 \caption{
   The RAR from weak lensing data (yellow diamonds), derived using our new deprojection formula Eq.~\eqref{eq:gobs_from_esd} and assuming our stellar and gas mass estimates described in Sec.~\ref{sec:masses}.
   The binned kinematic RAR from Ref.~\cite{Lelli2017b} is shown as gray circles.
   The error bars indicate the statistical uncertainty on the mean in each bin (not the galaxy-by-galaxy variation).
   The colored band indicates the systematic uncertainties of the lensing result from interpolation between the discrete bins and from extrapolation beyond the last data point, see Sec.~\ref{sec:newmethod}.
   The gray band indicates an additional systematic uncertainty of about $0.2\,\mathrm{dex}$ in stellar mass \citep{Brouwer2021}.
   For simplicity, we translate this $0.2\,\mathrm{dex}$ uncertainty into a $\sim 0.1\,\mathrm{dex}$ uncertainty on $\gobs$ using the fact that $\gobs$ scales as $\sqrt{\gbar}$.
   The dashed gray line shows the fitting function Eq.~\eqref{eq:nufunction} from Ref.~\cite{Lelli2017b}.
   The shaded region at $\gbar < 10^{-13}\,\mathrm{m/s}^2$ indicates where the isolation criterion may be less reliable according to the estimate from Ref.~\cite{Brouwer2021}.
   Our results from Sec.~\ref{sec:etgltg} suggest that LTGs may be sufficiently isolated down to $\gbar \approx 10^{-14}\,\mathrm{m/s}^2$.
   We shade this region where LTGs may still be reliable in a lighter color.
 }
 \label{fig:rar}
\end{figure}

\begin{table}
\caption{
  The weak-lensing radial acceleration relation from Fig.~\ref{fig:rar}.
  The kinematic data shown in Fig.~\ref{fig:rar} is available from Ref.~\cite{Lelli2017b}, see also \url{http://astroweb.cwru.edu/SPARC/}.
  The accelerations $\gbar$ and $\gobs$ are measured in $\mathrm{m/s}^2$. Uncertainties on $\gobs$ are converted to $\log_{10}$ space using linear error propagation.
  The systematic uncertainties listed here come from interpolating between discrete radial bins and from extrapolating beyond the last radial bin (see Sec.~\ref{sec:newmethod}, colored band in Fig.~\ref{fig:rar}).
  For brevity, the additional, fixed $0.1\,\mathrm{dex}$ systematic uncertainties on $\gobs$ from uncertainties in the stellar masses (shown as a gray band in Fig.~\ref{fig:rar}) are not listed here.
}
\label{table:rar}
\centering
\begin{tabular}{c c c c}
\\[-0.5em] %
\hline\hline
\\[-0.9em] %
$\log_{10} \gbar$&
$\log_{10} \gobs$&
$\sigma_{\log_{10} \gobs}^{\mathrm{statistical}}$ &
$\sigma_{\log_{10} \gobs}^{\mathrm{systematic}}$
\\
\hline
\\[-0.9em] %
\input{./plots/RAR-table.tex}
\hline
\end{tabular}
\end{table}

Fig.~\ref{fig:rar} and Table~\ref{table:rar} show the weak-lensing RAR obtained using our new method for converting excess surface densities to radial accelerations from Sec.~\ref{sec:newmethod} and our stellar and gas mass estimates from Sec.~\ref{sec:masses}.
The weak-lensing RAR smoothly continues the RAR obtained from kinematics \citep{Lelli2017b}.
We also show the fit function used in Ref.~\cite{Lelli2017b}, extrapolated to the smaller accelerations probed by weak lensing,
\begin{align}
\label{eq:nufunction}
g_{\mathrm{obs}} = \frac{g_{\mathrm{bar}}}{1 - e^{-\sqrt{g_{\mathrm{bar}}/a_0}}} \equiv g_{\mathrm{bar}} \, \nu\left(\frac{g_{\mathrm{bar}}}{a_0}\right) \,,
\end{align}
adopting $a_0 = 1.24\cdot 10^{-10}\,\mathrm{m}/\mathrm{s}^2$ \citep{McGaugh2011}.
This extrapolated fit function matches the weak lensing data well, except at the last few data points at $g_{\mathrm{bar}} \lesssim 10^{-14}\,\mathrm{m/s}^2$ where systematic uncertainties are large.
We further discuss these last few data points in Sec.~\ref{sec:discussion}.
In Appendix~\ref{sec:appendix:SIScompare}, we compare the RAR obtained using our new method to that obtained using the SIS method used in Ref.~\cite{Brouwer2021}.

The reason for the large systematic errors below $\gbar \approx 10^{-14}\,\mathrm{m/s}^2$ is the need to extrapolate beyond the last ESD data point, as explained in Sec.~\ref{sec:newmethod}.
Thus, in principle, we could likely get formally smaller systematic errors at $g_{\mathrm{bar}} \lesssim 10^{-14}\,\mathrm{m/s}^2$ by simply extending our analysis to values of $g_{\mathrm{bar}}$ below our current limit of $g_{\mathrm{bar}} = 10^{-15}\,\mathrm{m/s}^2$.
However, in practice, this is not very useful for two reasons.
First, observational systematics become more than just a small correction beyond a few $\mathrm{Mpc}$, see Appendix~\ref{sec:appendix:randomESD}.
If we extended our analysis to, say, $\gbar = 10^{-16}\,\mathrm{m/s}^2$, a typical lens galaxy with $M_b = 5 \cdot 10^{10}\,M_\odot$ would be probed up to $R \sim 8\,\mathrm{Mpc}$.
As described in Appendix~\ref{sec:appendix:errors}, we deal with such systematics by subtraction.
That procedure may, however, be suspect when these systematics are no longer just a small correction.
Second, the isolation criterion becomes less reliable at large radii.
Indeed, we have verified that the last few data points shown in Fig.~\ref{fig:rar} are sensitive to the precise choice of $R_{\mathrm{isol}}$ (see also Sec.~\ref{sec:etgltg}).
At even smaller values of $g_{\mathrm{bar}}$, such effects would become even more important.
Thus, the systematic uncertainties shown in Fig.~\ref{fig:rar} do indicate where systematic uncertainties become important, at least qualitatively.

The shaded region in Fig.~\ref{fig:rar} indicates where the isolation criterion may be less reliable.
This region corresponds to $\gbar < 10^{-13}\,\mathrm{m/s}^2$ where the isolation criterion from Ref.~\cite{Brouwer2021} may no longer be reliable according to their estimate.
We conservatively adapt this estimate despite using a stricter isolation criterion, namely $R_{\mathrm{isol}} = 4\,\mathrm{Mpc}/h_{70}$ instead of $R_{\mathrm{isol}} = 3\,\mathrm{Mpc}/h_{70}$.
Our results from Sec.~\ref{sec:etgltg} indicate that LTGs are sufficiently isolated at even smaller $\gbar$, down to $\gbar \approx 10^{-14}\,\mathrm{m/s}^2$.
It's only the ETGs that we are concerned about already at $\gbar \approx 10^{-13}\,\mathrm{m/s}^2$.
This is why we shade the region above $10^{-14}\,\mathrm{m/s}^2$, where LTGs may still be reliable, in a lighter color.

Our method for converting ESD profiles to radial accelerations is based on the deprojection formula Eq.~\eqref{eq:gobs_from_esd} which involves arbitrarily large radii.
Thus, if the isolation criterion fails at large radii (small $g_{\mathrm{bar}}$), this may in principle affect even the radial accelerations we infer at small radii (large $g_{\mathrm{bar}}$).
However, as we demonstrate in Appendix~\ref{sec:appendix:cutab14}, this is not a problem in practice.
We find almost identical results if we artificially cut off the ESD profiles that enter the integral formula Eq.~\eqref{eq:gobs_from_esd} at $g_{\mathrm{bar}} = 10^{-14}\,\mathrm{m/s}^2$.

\subsection{Similar RAR for early- and late-type galaxies}
\label{sec:etgltg}

\begin{figure}
 \centering
  \includegraphics[width=.8\textwidth]{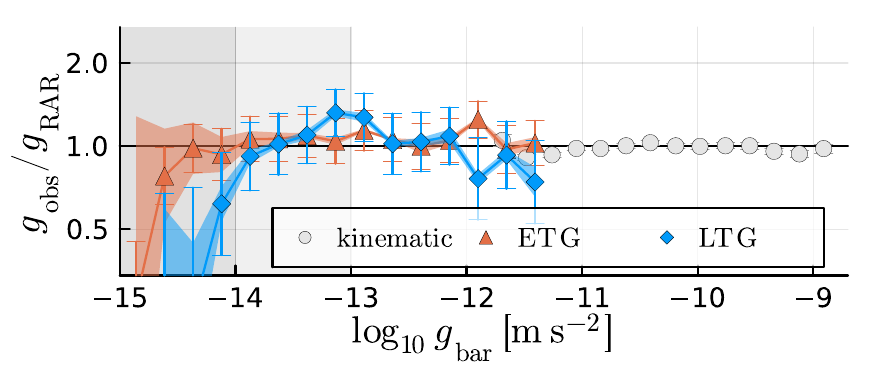}
  \caption{
  The acceleration implied by weak lensing for ETGs (red triangles) and LTGs (blue diamonds), relative to that of the fit function Eq.~\eqref{eq:nufunction} from Ref.~\cite{Lelli2017b}, here denoted by $g_{\mathrm{RAR}}$.
  The kinematic data from Ref.~\cite{Lelli2017b} is shown as gray circles.
  Error bars and bands are as in Fig.~\ref{fig:rar}, except we do not show the stellar mass systematic uncertainty for clarity.
  In contrast to Ref.~\cite{Brouwer2021}, we find good agreement between ETGs and LTGs down to about $\gbar = 10^{-14}\,\mathrm{m/s}^2$.
  The reason is that we use a stricter isolation criterion, $R_{\mathrm{isol}} = 4\,\mathrm{Mpc}/h_{70}$, and larger baryonic masses for ETGs (see Sec.~\ref{sec:masses}).
 }
 \label{fig:etgltg-Risol-4.0}
\end{figure}

In Ref.~\cite{Brouwer2021}, it is found that the RAR of ETGs deviates from that of LTGs.
In particular, according to Ref.~\cite{Brouwer2021}, LTGs follow the fit function Eq.~\eqref{eq:nufunction} of Ref.~\cite{Lelli2017b} even at extremely small accelerations $\gbar$, while ETGs deviate from it.
Here, we argue that -- with our stricter isolation criterion and our mass estimates from Sec.~\ref{sec:masses} -- weak-lensing data does not indicate a difference in the RAR of ETGs and LTGs.\footnote{
  An alternative scenario where hot gas increases the baryonic masses of ETGs by a factor of about two is also considered in Ref.~\cite{Brouwer2021}.
  Here, we confirm that the masses of ETGs may be related to the discrepancy found in Ref.~\cite{Brouwer2021}.
  In addition, we show that the isolation criterion plays a crucial role.
}
This fits with kinematic data which indicates that ETGs and LTGs follow the same RAR at accelerations $\gbar$ above $10^{-12}\,\mathrm{m/s}^2$ \citep{Lelli2017b,Shelest2020}.

Specifically, we find no significant difference between ETGs and LTGs down to $\gbar \approx 10^{-14}\,\mathrm{m/s}^2$, see Fig.~\ref{fig:etgltg-Risol-4.0}.
As we argue below, the difference to Ref.~\cite{Brouwer2021} is in the baryonic masses of ETGs (Sec.~\ref{sec:masses}) and the isolation criterion.
Following Ref.~\cite{Brouwer2021}, we here restrict the ETG and LTG subsamples to have the same stellar mass distribution (see the ``overlap'' shown in Fig.~\ref{fig:Mstar}) by randomly removing an appropriate number of galaxies at each stellar mass value.\footnote{
  We have verified that using different random numbers leads to similar results.
}

In Appendix~\ref{sec:appendix:BrouwerETGLTG} we show that we can reproduce the result of Ref.~\cite{Brouwer2021} if we adopt their baryonic masses and isolation criterion, i.e. if we adopt $R_{\mathrm{isol}} = 3\,\mathrm{Mpc}/h_{70}$, $Q_{\mathrm{ETG}} = 1.0$, and their gas mass estimate instead of our preferred values $R_{\mathrm{isol}} = 4\,\mathrm{Mpc}/h_{70}$, $Q_{\mathrm{ETG}} = 1.4$, and our gas mass estimate (see Sec.~\ref{sec:masses}).
Thus, the difference between our result and that of Ref.~\cite{Brouwer2021} is to be found in these differences in the masses and isolation criterion and not in the difference in the method for converting ESD profiles to radial accelerations (see Sec.~\ref{sec:newmethod}).

We further show in Appendix~\ref{sec:appendix:BrouwerETGLTG} that using our baryonic masses while keeping the isolation criterion of Ref.~\cite{Brouwer2021}, $R_{\mathrm{isol}} = 3\,\mathrm{Mpc}/h_{70}$, helps reduce the difference between ETGs and LTGs, but does not eliminate it.
Thus, both the isolation criterion and the baryonic masses of the lenses are important.
In the rest of this section, we will look into this more quantitatively.
We will show that ETGs and LTGs indeed agree reasonably well when using our default choices for the isolation criterion and the baryonic masses of the lenses.
In addition, we will argue that LTGs are sufficiently isolated down to about $\gbar \approx 10^{-14}\,\mathrm{m/s}^2$, i.e. further down than the validity limit of the isolation criterion from Ref.~\cite{Brouwer2021} according to their estimate, namely $\gbar \approx 10^{-13}\,\mathrm{m/s}^2$.
The same does not hold for ETGs, which are much more sensitive to the precise choice of isolation criterion.

\begin{figure}
 \centering
  \includegraphics[width=.8\textwidth]{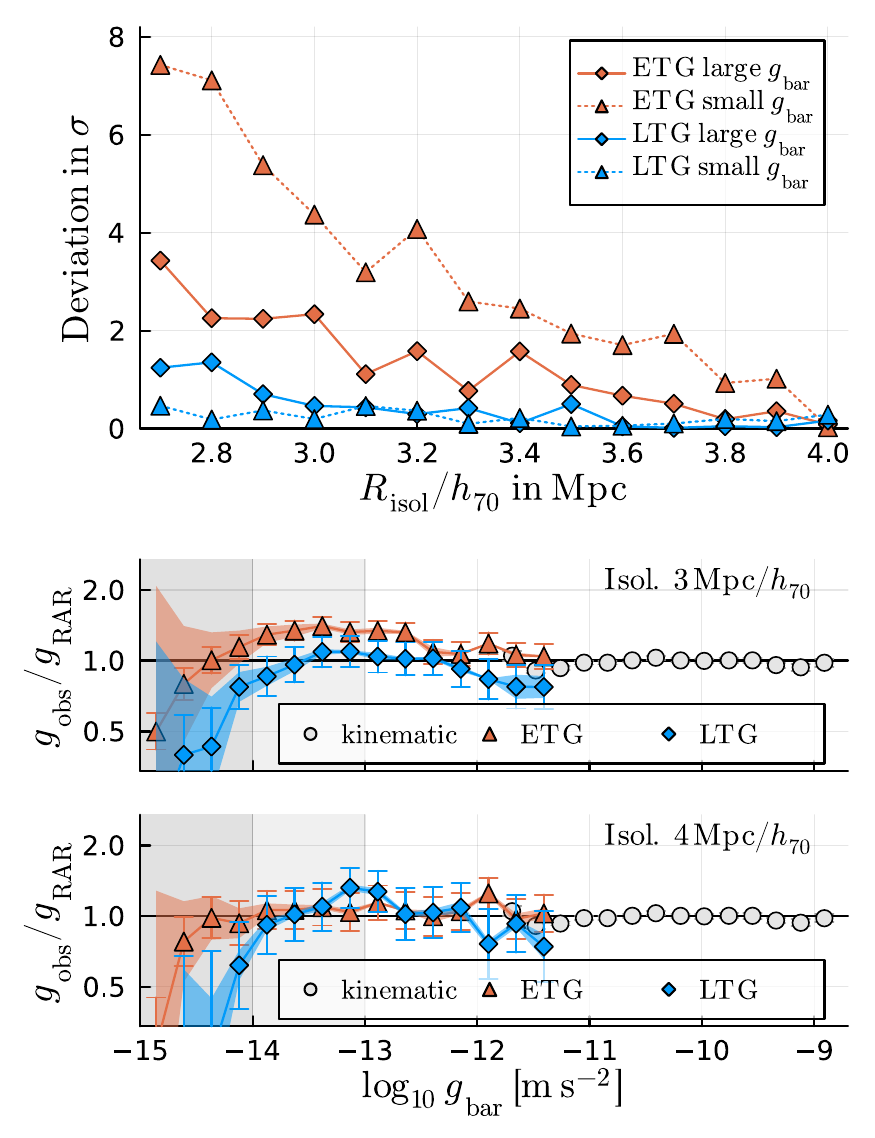}
 \caption{
   Top: The difference between the radial accelerations inferred from weak lensing and the RAR fitting function Eq.~\eqref{eq:nufunction}, measured in sigmas, as a function of how isolated the lenses are, quantified by $R_{\mathrm{isol}}$.
   We separately show the result for ETGs (red) and LTGs (blue) as well as for small (triangles with dashed lines) and large accelerations (diamonds with solid lines), see Eq.~\eqref{eq:defsmalllargegbar}.
   LTGs are mostly unaffected by making the isolation criterion stricter.
   In contrast, ETGs do depend on $R_{\mathrm{isol}}$, but, with increasing $R_{\mathrm{isol}}$, tend towards what the fitting function Eq.~\eqref{eq:nufunction} predicts.
   Middle and bottom: The actual accelerations behind these sigma numbers for $R_{\mathrm{isol}} = 3\,\mathrm{Mpc}/h_{70}$ and $R_{\mathrm{isol}} = 4\,\mathrm{Mpc}/h_{70}$.
 }
 \label{fig:etgltg-sigmas}
\end{figure}

To understand the effect of the isolation criterion, we try various values of $R_{\mathrm{isol}}$, see Fig.~\ref{fig:etgltg-sigmas}.
Following Ref.~\cite{Brouwer2021} we quantify the difference between the lensing-inferred radial accelerations and those implied by the fit function Eq.~\eqref{eq:nufunction} by first calculating a $\chi^2$ value and then converting this into a number of $\sigma$.
For simplicity, we treat different radii as independent, i.e. we leave out the small off-diagonal elements of the covariance matrix \citep{Brouwer2021}.
Concretely,
\begin{align}
 \label{eq:chi2}
 \chi^2 \equiv \sum_i \frac{
   (g_{\mathrm{obs},\mathrm{lensing},i} - g_{\mathrm{bar},i} \, \nu(g_{\mathrm{bar},i}/a_0))^2
  }{
    \sigma^2_{g_{\mathrm{obs},i}}|_{\mathrm{statistical}}
  } \,,
\end{align}
where $i$ runs over the $g_{\mathrm{bar}}$ bins, $g_{\mathrm{obs},\mathrm{lensing}}$ is the lensing-inferred radial acceleration, and $\left.\sigma^2_{g_{\mathrm{obs}}}\right|_{\mathrm{statistical}}$ is the statistical uncertainty of $g_{\mathrm{obs},\mathrm{lensing}}$.
Appendix~\ref{sec:appendix:BrouwerETGLTG} suggests that the difference between ETGs and LTGs is larger at relatively small accelerations and smaller at relatively large accelerations.
To keep track of this, we separately consider large and small accelerations,
\begin{align}
\label{eq:defsmalllargegbar}
\begin{aligned}
 &\mathrm{small}\;g_{\mathrm{bar}}: &10^{-14}\,\mathrm{m/s}^2 - 10^{-13}\,\mathrm{m/s}^2 \,, \\
 &\mathrm{large}\;g_{\mathrm{bar}}: &10^{-13}\,\mathrm{m/s}^2 - 10^{-11}\,\mathrm{m/s}^2 \,.
\end{aligned}
\end{align}
We do not consider data below $g_{\mathrm{bar}} = 10^{-14}\,\mathrm{m/s}^2$ because there the systematic uncertainties become important, while $\chi^2$ takes into account only the statistical uncertainties.

We can now discuss the effect of $R_{\mathrm{isol}}$ shown in Fig.~\ref{fig:etgltg-sigmas}.
We first note that LTGs are barely affected by changes in $R_{\mathrm{isol}}$.
This holds for both small and large accelerations $g_{\mathrm{bar}}$.
LTGs follow the fit function Eq.~\eqref{eq:nufunction} quite well down to $g_{\mathrm{bar}} \approx 10^{-14}\,\mathrm{m/s}^2$.
That there is no dependence on $R_{\mathrm{isol}}$ indicates that the LTG sample is sufficiently isolated down to $g_{\mathrm{bar}} \approx 10^{-14}\,\mathrm{m/s}^2$, as we will discuss in more detail below.
At even smaller $g_{\mathrm{bar}}$ -- in the very tail of our lensing RAR -- systematic uncertainties become important and even the LTGs are sensitive to $R_{\mathrm{isol}}$, so one should be cautious about this tail.

The situation is different for ETGs.
With $R_{\mathrm{isol}} = 3\,\mathrm{Mpc}/h_{70}$, they deviate significantly from the fit function Eq.~\eqref{eq:nufunction}, especially at relatively small accelerations $g_{\mathrm{bar}}$, but this deviation becomes smaller as we increase $R_{\mathrm{isol}}$.
That the trend with $R_{\mathrm{isol}}$ is stronger for small accelerations fits with the fact that the isolation criterion is expected to be less reliable at large radii \citep{Brouwer2021}.
In any case, this result shows that for ETGs one must be careful with the isolation criterion.
A too small $R_{\mathrm{isol}}$ may result in an artificially increased signal, likely due to the two-halo term \citep{Brouwer2021}.
This is why our default choice is $R_{\mathrm{isol}} = 4\,\mathrm{Mpc}/h_{70}$, i.e. a stricter isolation criterion compared to Ref.~\cite{Brouwer2021}.

That ETGs are more sensitive to the isolation criterion than LTGs makes sense from the perspective of the morphology-density relation, i.e. considering that ETGs are more clustered than LTGs \citep{Dressler1980}.
That is, if the isolation criterion fails, we would expect it to fail earlier for ETGs.
This is consistent with what we see here.

Compared to Ref.~\cite{Brouwer2021}, we here consider a stricter isolation criterion, i.e. we make $R_{\mathrm{isol}}$ larger than $3\,\mathrm{Mpc}/h_{70}$.
As a result, the lens sample becomes smaller and the statistical uncertainties increase.
Thus, one may worry that the difference between LTGs and ETGs becomes statistically insignificant simply because of the larger uncertainties.
In Appendix~\ref{sec:appendix:rescalederrors}, we demonstrate that this is not the case; we find the same qualitative behavior if we artificially keep the error bars fixed at the values they have for $R_{\mathrm{isol}} = 3\,\mathrm{Mpc}/h_{70}$.

Fig.~\ref{fig:etgltg-sigmas} also shows that ETGs do not flatten-out as a function of $R_{\mathrm{isol}}$ in the same way LTGs do.
This is important because, if our results depend on $R_{\mathrm{isol}}$, this means we are not (just) measuring intrinsic properties of the lens galaxies but (also) something about their environment.
In contrast, when the trend flattens out -- as for the LTGs -- we are plausibly measuring an intrinsic property of the galaxies.\footnote{
  Strictly speaking, one may find a flattening-out at $0\,\sigma$ even with a non-isolated sample simply because the error bars increase with $R_{\mathrm{isol}}$.
  However, Appendix~\ref{sec:appendix:rescalederrors} shows that we find the same qualitative trends with $R_{\mathrm{isol}}$ even when we artificially keep the size of the error bars fixed (until fluctuations take over at large $R_{\mathrm{isol}}$, as we discuss below).
}
Thus, ETGs -- in contrast to LTGs -- may not be sufficiently isolated down to $\gbar \approx 10^{-14}\,\mathrm{m/s}^2$.
ETGs look more reliable down to $10^{-13}\,\mathrm{m/s}^2$, which matches where the isolation criterion from Ref.~\cite{Brouwer2021} is reliable according to their estimate.

It would be interesting to see if the trend of ETGs flattens out at $R_{\mathrm{isol}}$ larger than what we show in Fig.~\ref{fig:etgltg-sigmas}, i.e. at $R_{\mathrm{isol}} > 4\,\mathrm{Mpc}/h_{70}$.
We have verified that -- within the statistical uncertainties -- it does.
However, beyond $R_{\mathrm{isol}} = 4\,\mathrm{Mpc}/h_{70}$, statistical fluctuations due to the small sample size take over if we artificially keep the error bars fixed at the values they have for $R_{\mathrm{isol}} = 3\,\mathrm{Mpc}/h_{70}$.
So at $R_{\mathrm{isol}} > 4\,\mathrm{Mpc}/h_{70}$, we cannot be sure if the trend flattens out just due to the larger error bars (i.e. the argument from Appendix~\ref{sec:appendix:rescalederrors} no longer works).

We note that, at the strictest isolation criterion shown in Fig.~\ref{fig:etgltg-sigmas}, $R_{\mathrm{isol}} = 4\,\mathrm{Mpc}/h_{70}$, ETGs match the fit function Eq.~\eqref{eq:nufunction} almost perfectly down to $\gbar \approx 10^{-14}\,\mathrm{m/s}^2$.
However, since we are not seeing the trend flatten out, this almost perfect match may be a statistical fluctuation.
This leaves open the possibility that other $M_\ast/L$ values for ETGs, i.e. other values of $Q_{\mathrm{ETG}}$, may provide a better fit.
We have tried various values and find that $Q_{\mathrm{ETG}} = 1.8$ works well.\footnote{
  Alternatively, a similar result can be obtained by keeping the $M_\ast/L$ fixed and adjusting the hot gas mass estimate.
  This is because, as already mentioned in Sec.~\ref{sec:masses}, our results are mainly sensitive to the total baryonic mass.
}
However, because we cannot be sure that ETGs are sufficiently isolated, these numbers should not be overinterpreted.

To sum up, our LTG sample seems to be sufficiently isolated to derive a robust lensing RAR down to $\gbar \approx 10^{-14}\,\mathrm{m/s}^2$.
Indeed, both ETGs and LTGs follow the fitting function Eq.~\eqref{eq:nufunction} quite well down to $\gbar \approx 10^{-14}\,\mathrm{m/s}^2$ when we impose a sufficiently strict interpolation criterion and use our mass estimate from Sec.~\ref{sec:masses}.
However, the results for ETGs may not be reliable all the way down to $\gbar \approx 10^{-14}\,\mathrm{m/s}^2$ because these depend quite a bit on the details of the isolation criterion that one adopts.
In any case, we find no significant difference between the RAR for LTGs and ETGs.

\subsection{Mass bins}
\label{sec:massbins}

\begin{figure}
 \centering
  \includegraphics[width=\textwidth]{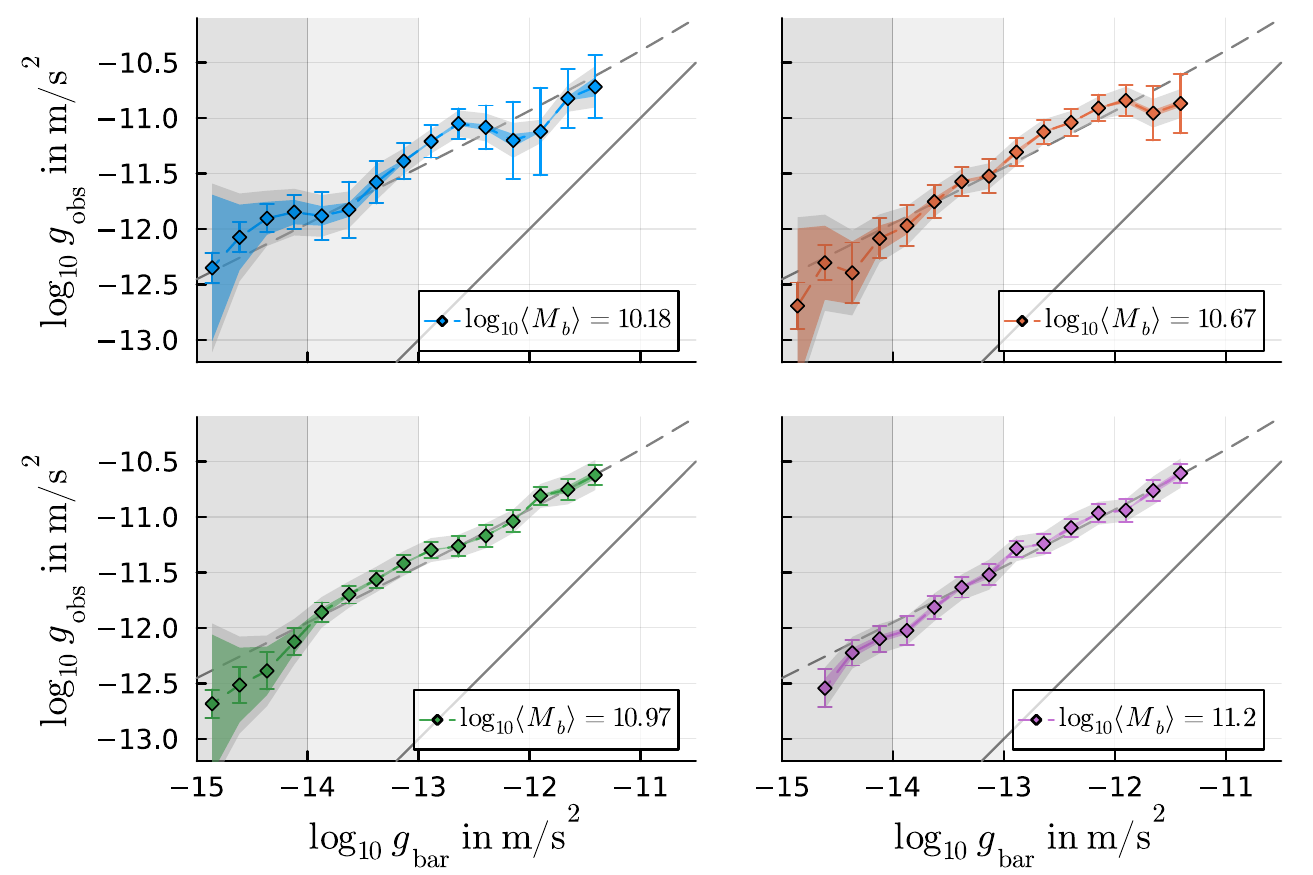}
  \caption{
    The RAR implied by weak lensing for four baryonic mass bins with bin edges $\log_{10} M_b/M_\odot = [9.0, 10.5, 10.8, 11.1, 11.5]$ assuming our mass estimates from Sec.~\ref{sec:masses}.
    The error bars and bands are as in Fig.~\ref{fig:rar}.
    All mass bins generally agree with the RAR fitting function Eq.~\eqref{eq:nufunction} from Ref.~\cite{Lelli2017b}.
    In the smallest mass bin, the radial accelerations $g_{\mathrm{obs}}$ tend to fall above this fitting function's prediction more than for the other mass bins.
    This may be because the isolation criterion is less reliable for small masses.
    The last data point in the highest-mass bin is not shown because the inferred $g_{\mathrm{obs}}$ is negative there.
  }
  \label{fig:RAR-Mbbins-ETG-1.4}
\end{figure}

Fig.~\ref{fig:RAR-Mbbins-ETG-1.4} shows the RAR separately for four baryonic mass bins, with $\log_{10} M_b/M_\odot$ bin edges $[9.0, 10.5, 10.8, 11.1, 11.5]$.
All mass bins generally show similar behavior, following the fit function Eq.~\eqref{eq:nufunction} down to $g_{\mathrm{bar}} \approx 10^{-14}\,\mathrm{m/s}^2$.
This is especially true for larger baryonic masses.
For smaller masses, there are larger deviations.
A part of this is likely due to statistical fluctuations.
Indeed, the sample is smaller for smaller masses, as indicated by the larger error bars in Fig.~\ref{fig:RAR-Mbbins-ETG-1.4}.
Still, there may be a small systematic shift to larger accelerations $\gobs$.

If there is such a systematic shift, a possible explanation is a failure of the isolation criterion:
In contrast to larger galaxies, smaller galaxies can be affected even by a relatively small neighbor.
This is exacerbated by the $m = 20$ magnitude limit of the KiDS bright sample \citep{Brouwer2021,Bilicki2021} which means that many of the neighbors that affect small lenses are not detected.

The effect of the magnitude limit was discussed in Appendix~A of Ref.~\cite{Brouwer2021} who estimated that the effect on their results is small.
However, only the effect on the whole lens sample was considered in Ref.~\cite{Brouwer2021}, not the effect on a small-mass subsample.
Indeed, small-mass galaxies are only a small part of the whole lens sample so, overall, they have only a small effect.
If one splits the sample by mass, however, the small-mass bins may be affected by this systematic effect.
Thus, we advise some caution when considering small-mass subsamples.\footnote{
  That the isolation criterion may be less reliable for small masses is also the reason we impose an isolation criterion in position space rather than in $g_{\mathrm{bar}}$ space.
  More specifically, many of our results are about quantities stacked in $g_{\mathrm{bar}}$ space.
  So a natural choice would be to demand that galaxies are isolated down to some small value $g_{\mathrm{bar},\mathrm{isol}}$ of $g_{\mathrm{bar}}$ rather than out to some large value $R_{\mathrm{isol}}$ of $R$.
  However, demanding isolation in position space has the advantage that -- at a given $g_{\mathrm{bar}}$ -- it is automatically more strict for small-mass galaxies because of the factor of $M_b$ in $g_{\mathrm{bar},\mathrm{isol}} = G M_b/R_{\mathrm{isol}}^2$.
}

We have explored lifting our 
upper cutoff in stellar mass and considered a fifth mass bin including lenses up to $\log_{10} M_b/M_\odot = 11.9$. We find that the overall shape of the lensing RAR remains the same but with a slight offset of $g_{\mathrm{obs}}$ towards larger values.
One possible reason is that the isolation criterion fails for such high-mass galaxies because they have more satellites. Indeed, this is the reason the stellar mass cut was originally introduced in Ref.~\cite{Brouwer2021}.  However, the offset we find is basically constant across all $g_{\mathrm{bar}}$ values which is in conflict with the expectation that the isolation criterion should break down gradually towards larger radii. In contrast, a roughly constant offset is expected if we underestimate the $M_\ast/L$ for high-mass galaxies.
The high-mass end of our sample is dominated by ETGs and the offset we are seeing in their RAR may be an indication of 
a mass-dependent $M_\ast/L$, which is a possibility we did not consider %
when calculating our $Q_{\mathrm{ETG}}$. A value of $Q_{\mathrm{ETG}} = 1.8$ roughly compensates the offset in $g_{\mathrm{obs}}$, implying that our baseline stellar masses for the most massive ETGs might need to be increased by a factor $1.8/1.4=1.29$. The required change seems plausible in terms of stellar population models \cite{Schombert2019}, especially given the possible need for a mass-dependent IMF \cite{Cappellari2012,Conroy2012,Conroy2013}, a topic  %
beyond the scope of this paper.

\section{Discussion}
\label{sec:discussion}

In Sec.~\ref{sec:rar} we saw that the RAR shows signs of a downturn in the last few data points at $g_{\mathrm{bar}} \lesssim 10^{-14}\,\mathrm{m/s}^2$.
Due to the systematic uncertainties discussed there, one should be careful not to read too much into these last few data points.
Still, if the downturn is real, it would be an interesting subject of further study.

For example, in the context of particle dark matter, a downturn of the RAR is expected due to the finite size of the dark matter halo.
In fact, such a downturn should happen already well before $g_{\mathrm{bar}} = 10^{-14}\,\mathrm{m/s}^2$ \citep[e.g.][]{Li2022,Mercado2023}.
Thus, the extended part of the weak-lensing RAR before the potential downturn is likely to provide strong constraints on particle dark matter models.
Indeed, the density of an NFW tail falls off as $1/r^3$ while, before the potential downturn, the weak-lensing RAR requires a $1/r^2$ profile.

In the context of modified gravity, a downturn at large radii may be expected for different reasons, related to the lenses no longer being isolated.
Most relevant here are models that reduce to Modified Newtonian Dynamics \citep[MOND,][]{Milgrom1983a,Milgrom1983b,Milgrom1983c} in the non-relativistic limit.
Many of these have the attractive property that they naturally explain why the weak-lensing RAR follows the MOND-inspired fit function Eq.~\eqref{eq:nufunction} so well \citep{Milgrom2013,Famaey2012}.
Importantly, many of these models also feature a so-called External Field Effect (EFE) which leads to a downturn at large radii where galaxies can no longer be treated as isolated \citep{Famaey2012,Chae2022b}.\footnote{
  In some models that combine features of MOND and features of particle dark matter -- such as Aether Scalar  Tensor Theory \citep{Skordis2020} -- there may be yet another type of downturn, unrelated to non-isolation.
  This was considered in Ref.~\cite{Mistele2023} and may be in tension with weak-lensing data.
}
This may explain the possible downturn of the lensing RAR at $g_{\mathrm{bar}} \lesssim 10^{-14}\,\mathrm{m/s}^2$.

Unfortunately, quantitatively testing the EFE is complicated.
One reason is that the EFE works differently in different models of MOND.
Another is that one would actually expect two competing effects in such models.
One is the EFE which one expects to lead to a downturn.
The other is a version of the so-called two-halo term which tends to lead to an upturn instead.
Thus, one would have to quantitatively work out the net effect of these two competing effects.

A final complication is relevant even beyond testing modified gravity models.
Namely, the isolation criterion we apply to the lens galaxies is ultimately based on photometric redshifts.
These redshifts are estimated to have an uncertainty of $\sigma_z = 0.02 \, (1 + z)$ \citep{Brouwer2021}.
Since the relevant redshifts $z$ are in the range $0.1$ to $0.5$, these uncertainties translate into a distance uncertainty much larger than the value $R_{\mathrm{isol}} = 4\,\mathrm{Mpc}/h_{70}$ we use for our isolation criterion.
This complicates any quantitative analysis of the EFE and other effects related to non-isolation, and is a reminder of the importance of accurate spectroscopic redshifts.

More generally, this means our lens sample is unlikely to be as isolated as the formal definition of $R_{\mathrm{isol}}$ suggests (no other galaxies with at least $10\%$ of a lenses' mass within $R_{\mathrm{isol}}$).
We nevertheless trust our results for the following reasons.
First, the isolation criterion from Ref.~\cite{Brouwer2021} was validated against simulations and found to be trustworthy down to about $\gbar = 10^{-13}\,\mathrm{m/s}^2$.
Since we use a stricter isolation criterion, our results should also be trustworthy down to at least $\gbar = 10^{-13}\,\mathrm{m/s}^2$.
Second, we have seen in Sec.~\ref{sec:etgltg} that, down to $\gbar \approx 10^{-14}\,\mathrm{m/s}^2$, the RAR for LTGs is almost independent of the value of $R_{\mathrm{isol}}$.
This indicates that LTGs are sufficiently isolated down to $g_{\mathrm{bar}} \approx 10^{-14}\,\mathrm{m/s}^2$.
ETGs, in contrast, are quite sensitive to the precise value of $R_{\mathrm{isol}}$.
This makes sense considering that ETGs are more clustered \citep{Dressler1980}, but means we cannot trust our results for ETGs quite as far down in $\gbar$.
In particular, as discussed above, we trust our results for ETGs down to $\gbar \approx 10^{-13}\,\mathrm{m/s}^2$ and those for LTGs down to $\gbar \approx 10^{-14}\,\mathrm{m/s}^2$.

At even smaller accelerations, $\gbar \lesssim 10^{-14}\,\mathrm{m/s}^2$, both ETGs and LTGs are unlikely to be sufficiently isolated.
This fits well with what we suggested above in the context of modified gravity models;
namely that the possible downturn of the lensing RAR at $\gbar \lesssim 10^{-14}\,\mathrm{m/s}^2$ may be due to the isolation criterion breaking down, and not due to an intrinsic property of isolated galaxies.
Indeed, there is a limit to how isolated galaxies can be due to the large-scale structure of the universe \citep{Chae2021b}.
Investigating this in more detail is left for future work, as is the analysis of larger spectroscopic surveys as data improve.

\section{Conclusion}
\label{sec:conclusion}

We have combined weak-lensing and kinematic data to construct the RAR over a large dynamic range in acceleration.
We have estimated the lens galaxies' stellar and gas masses consistently with previous kinematic determinations of the RAR.
We have further employed a new deprojection formula that converts excess surface densities to radial accelerations for spherically symmetric lenses.

We find that the RAR inferred from weak lensing smoothly continues the RAR from kinematic data by about $2.5\,\mathrm{dex}$, implying a dark matter density profile that scales as $1/r^2$ out to large radii.
In contrast to previous studies we find no significant difference between the RAR for ETGs and LTGs.
This is partly due to our somewhat larger baryonic masses for ETGs and partly due to the fact that we impose a stricter isolation criterion.

At the last few data points -- where systematic uncertainties are important -- we find hints of a downturn of the RAR.
We speculate that this may be related to a failure of the isolation criterion rather than an intrinsic property of the lenses.
We will further investigate this in future work.

\acknowledgments

TM thanks Margot Brouwer for patiently answering many questions and Sabine Hossenfelder for helpful discussions.
TM is grateful to Christian Holm and the ICP Stuttgart for their hospitality.
This work was supported by the DFG (German Research Foundation) – 514562826.
PL is supported by the Alexander von Humboldt Foundation.
Based on observations made with ESO Telescopes at the La Silla Paranal Observatory under programme IDs 177.A-3016, 177.A-3017, 177.A-3018 and 179.A-2004, and on data products produced by the KiDS consortium. The KiDS production team acknowledges support from: Deutsche Forschungsgemeinschaft, ERC, NOVA and NWO-M grants; Target; the University of Padova, and the University Federico II (Naples).

\appendix

\section{Derivation of the deprojection formula Eq.~\eqref{eq:gobs_from_esd}}
\label{sec:appendix:derivation}

To obtain the radial acceleration $g_{\mathrm{obs}}$ from the excess surface density $\Delta \Sigma$ for a single isolated galaxy, we proceed in two steps.
First, we find the projected surface density $\Sigma$ from the excess surface density $\Delta \Sigma$.
Then, we plug the result in a well-known deprojection formula from Ref.~\cite{Kent1986} which gives the radial acceleration $g_{\mathrm{obs}}$ in terms of the surface density $\Sigma$.
The resulting expression can be simplified to give Eq.~\eqref{eq:gobs_from_esd}.

We begin with the relation
\begin{align}
 \partial_R \left(\Delta \Sigma(R) R^2 \right) = - R^2 (\partial_R \Sigma(R)) \,.
\end{align}
This follows directly from the definition Eq.~\eqref{eq:esd} of $\Delta \Sigma$ in terms of $\Sigma$.
We can divide both sides by $R^2$ and then integrate from $R$ to $\infty$ to find
\begin{align}
\begin{split}
 \Sigma(R) - \Sigma(\infty)
 &= \int_R^{\infty} dR' \frac{\partial_{R'}(\Delta \Sigma(R') R'^2)}{R'^2}\\
 &=  \Delta \Sigma(\infty) - \Delta \Sigma(R) + \int_R^{\infty} dR' \frac{2 \Delta \Sigma(R')}{R'} \,.
\end{split}
\end{align}
In the following, we assume that the absolute value of the density, $|\rho|$, asymptotically falls off at least as fast as $r^{-n}$ for some $n > 1$.
This implies that the surface density $\Sigma(R)$ is finite and that both $\Sigma(R)$ and $\Delta \Sigma(R)$ fall off at least as fast as $R^{1-n}$.
In particular, $\Sigma$ and $\Delta \Sigma$ vanish at infinity,
\begin{align}
 \Sigma(\infty) = 0 \,, \quad \Delta \Sigma(\infty) = 0 \,.
\end{align}
Thus, we have
\begin{align}
 \label{eq:sd_from_esd}
 \Sigma(R) =  - \Delta \Sigma(R) + \int_R^{\infty} dR' \frac{2 \Delta \Sigma(R')}{R'} \,.
\end{align}
This allows to calculate $\Sigma$ from $\Delta \Sigma$ and completes the first step.

The central ingredient for the second step is the following relation between $g_{\mathrm{obs}}$ and $\Sigma$ from Ref.~\cite{Kent1986} which holds in spherical symmetry\footnote{
  We have rederived this result, making sure it does not require assumptions on $\rho$ that are stronger than what we assume here.
  In particular, we have verified that it does not require positive densities.
}
\begin{align}
 \label{eq:VcKent}
 g_{\mathrm{obs}}(r) =  \frac{G}{r^2} \left(I_1 + I_2\right) \,,
\end{align}
with the spherical radius $r$ and the integrals
\begin{align}
 I_1 &\equiv \int_0^r dR \, 2\pi R \, \Sigma(R) \,, \\
 I_2 &\equiv 4 \int_r^{\infty} dR\, R \Sigma(R) \left(\arcsin\left(\frac{r}{R}\right) - \frac{r}{\sqrt{R^2-r^2}}\right) \,.
\end{align}
Note that the first of these integrals appears directly in the definition Eq.~\eqref{eq:esd} of $\Delta \Sigma$ in terms of $\Sigma$.
Thus, this first integral is just $\pi r^2(\Delta \Sigma(r) + \Sigma(r))$.
Using Eq.~\eqref{eq:sd_from_esd} then gives
\begin{align}
 \label{eq:I1}
 I_1 = \pi r^2 ( \Delta \Sigma(r) + \Sigma(r) ) = 2 \pi r^2 \int_r^\infty dR \frac{\Delta \Sigma(R)}{R} \,.
\end{align}
This leaves the second integral.
Plugging in our result Eq.~\eqref{eq:sd_from_esd} gives
\begin{align}
\label{eq:I2}
\begin{split}
    I_2 =
      &-4 r \int_r^{\infty} dR\, \Delta\Sigma(R) f(R/r)  \\
      &+4 r \int_r^{\infty} dR\, f(R/r) \int_R^\infty dR' \frac{2 \Delta \Sigma(R')}{R'} \\
      \equiv &I_{21} + I_{22} \,,
\end{split}
\end{align}
where we introduced the short-hand notation
\begin{align}
 f(x) \equiv x \left( \arcsin\left(\frac{1}{x}\right) - \frac{1}{\sqrt{x^2-1}} \right) \,.
\end{align}
By plugging the expressions Eq.~\eqref{eq:I1} and Eq.~\eqref{eq:I2} for $I_1$ and $I_2$ into Eq.~\eqref{eq:VcKent} we obtain an expression for $g_{\mathrm{obs}}$ in terms of $\Delta \Sigma$.
It remains to simplify this expression.

Our first simplification step is to reduce the double integral $I_{22}$ to a single integral.
To facilitate this, we change the order of integration,
\begin{align}
\begin{split}
 I_{22}
 &= 4 r \int_r^{\infty} dR\, f(R/r) \int_R^\infty dR' \frac{2 \Delta \Sigma(R')}{R'} \\
 &= 4 r \int_r^{\infty} dR\, f(R/r) \int_0^\infty dR' \Theta(R' - R) \frac{2 \Delta \Sigma(R')}{R'} \\
 &= 4 r \int_0^\infty dR' \frac{2 \Delta \Sigma(R')}{R'} \int_r^\infty dR \, \Theta(R' - R) f(R/r) \,,
\end{split}
\end{align}
where $\Theta$ denotes the Heaviside step function.
The lower integration boundary of the $r$ integral and the $\Theta$ function ensure that we always have $r < R < R'$.
This implies we can set the lower integration boundary of the $R'$ integral to $r$,
\begin{align}
\begin{split}
 I_{22}
 &= 4 r \int_r^\infty dR' \frac{2 \Delta \Sigma(R')}{R'} \int_r^\infty dR \, \Theta(R' - R) f(R/r) \\
 &= 4 r \int_r^\infty dR' \frac{2 \Delta \Sigma(R')}{R'} \int_r^{R'} dR \, f(R/r) \\
 &= 4 r^2 \int_r^\infty dR' \frac{\Delta \Sigma(R')}{R'} \int_1^{R'/r} dx \, 2f(x) \,.
\end{split}
\end{align}
The $x$ integral can be done by Mathematica \citep{Mathematica13},
\begin{align}
 I_{22} = 4 r^2 \int_r^\infty dR \frac{\Delta \Sigma(R)}{R} \left(
   -\frac{\pi}{2}
   -\sqrt{\left(\frac{R}{r}\right)^2-1}
   +\left(\frac{R}{r}\right)^2 \arcsin\left(\frac{r}{R}\right)
  \right) \,.
\end{align}

All integrals can now be combined into a single integral with integrand proportional to $\Delta \Sigma$.
We have,
\begin{align}
 I_1 + I_{21} + I_{22} =  \int_r^\infty dR \Delta \Sigma(R) \, h(R) \,,
\end{align}
with the auxiliary function $h(R)$ given by
\begin{align}
\begin{split}
h(R) \equiv
   \frac{2 \pi r^2}{R}
   &- 4r \frac{R}{r} \left[ \arcsin\left(\frac{r}{R}\right) - \frac{1}{\sqrt{\left(\frac{R}{r}\right)^2 - 1}} \right]\\
   &+ 4r^2 \frac{1}{R} \left[
    -\frac{\pi}{2}
    -\sqrt{\left(\frac{R}{r}\right)^2-1}
    +\left(\frac{R}{r}\right)^2 \arcsin\left(\frac{r}{R}\right)
   \right] \,.
\end{split}
\end{align}
Some of these terms cancel and we end up with
\begin{align}
g_{\mathrm{obs}}(r) = \frac{4G}{r} \int_r^\infty dR \, \Delta \Sigma(R) \left[\frac{1}{\sqrt{1 - \left(\frac{r}{R}\right)^2}} - \sqrt{1 - \left(\frac{r}{R}\right)^2}\right] \,.
\end{align}
Our final simplification step is to substitute
\begin{align}
 \frac{r}{R} \equiv \sin \theta \,,
\end{align}
with $\theta$ in the interval $[0, \pi/2]$ which finally gives
\begin{align}
g_{\mathrm{obs}}(r) = 4G \int_0^{\pi/2} d\theta \, \Delta \Sigma\left(\frac{r}{\sin \theta}\right) \,.
\end{align}

\section{RAR from stacking}
\label{sec:appendix:newmethod}

\subsection{Stacking a large number of galaxies}
\label{sec:appendix:newmethod:stacked}

In Sec.~\ref{sec:newmethod:individual}, we have considered individual galaxies.
However, the weak-lensing signal for any individual lens galaxy is small.
Thus, one usually considers the stacked signal from a large sample of lens galaxies.

The stacked ESD profile used by Ref.~\cite{Brouwer2021} can be written in the form
\begin{align}
 \label{eq:esd_stacked}
 \Delta \Sigma^{\mathrm{stacked}}(R) = N^{-1}(R) \sum_l w_l(R) \, \Delta \Sigma_l(R) \,,
\end{align}
where the sum runs over the lens galaxies $l$, the $w_l(R)$ are unnormalized weights, $N(R) = \sum_l w_l(R)$ is a normalization factor, and $\Delta \Sigma_l(R)$ is an estimate for the ESD profile of the lens $l$ at the projected radius $R$.
We will explain how $w_l(R)$ and $\Delta \Sigma_l(R)$ are related to actual weak-lensing data below in Appendix~\ref{sec:appendix:newmethod:observational}.

But first we discuss how to go from ESD profiles to accelerations with such stacked data.
For this, we would like to use the exact deprojection formula Eq.~\eqref{eq:gobs_from_esd}.
The question is how to go from an individual lens to a large sample of stacked lenses, i.e. how to define a reasonable stacked radial acceleration $g_{\mathrm{obs}}^{\mathrm{stacked}}$,
\begin{align}
 g_{\mathrm{obs}}^{\mathrm{stacked}} = \, ? \,.
\end{align}
Note that the stacked ESD profile $\Delta \Sigma^{\mathrm{stacked}}$ is, at each radius $R$, a weighted average of the ESD profiles of the individual lenses.
So it has a straightforward physical interpretation as an average ESD profile.
We would like to have a similarly straightforward interpretation for the stacked acceleration $g_{\mathrm{obs}}^{\mathrm{stacked}}$.

Unfortunately, the simplest idea one might have for a stacked radial acceleration does not  in general have such a straightforward interpretation.
Namely, one might want to simply apply Eq.~\eqref{eq:gobs_from_esd} to the stacked ESD profile $\Delta \Sigma^{\mathrm{stacked}}$ (i.e. first stack, then deproject),
\begin{align}
 \label{eq:gobs_from_stacked}
 g_{\mathrm{obs}}^{\mathrm{stacked}}(R) \stackrel{?}{=} 4G \int_0^{\pi/2} d \theta \Delta \Sigma^{\mathrm{stacked}}\left(\frac{R}{\sin \theta}\right) \quad (\mathrm{not\;ideal}) \,.
\end{align}
But  this is \emph{not} a weighted average of the radial accelerations of the individual galaxies.
The reason is that the weights $w_l(R)$ in the stacked ESD profile depend on $R$.
Indeed, if we write out this definition in full, we have
\begin{align}
\label{eq:bad_stacked_written_out}
4G \int_0^{\pi/2} d\theta \, N^{-1}\left(\frac{R}{\sin \theta}\right) \sum_l w_l\left(\frac{R}{\sin \theta}\right) \Delta \Sigma_l\left(\frac{R}{\sin \theta}\right) \,.
\end{align}
In contrast, any weighted average of the radial accelerations of the individual lenses can be written in the form,
\begin{align}
\label{eq:good_stacked_general}
 g_{\mathrm{obs}}^{\mathrm{averaged}}(R) \equiv 4G \bar{N}^{-1}(R) \sum_l \bar{w}_l(R) \int_0^{\pi/2} d\theta \Delta \Sigma_l\left(\frac{R}{\sin \theta}\right) \,,
\end{align}
for some weights $\bar{w}_l(R)$ with normalization factor $\bar{N} = \sum_l \bar{w}_l(R)$.
Eq.~\eqref{eq:bad_stacked_written_out} is of this form only when the combination $N^{-1}(R) \, w_l(R)$ does not depend on $R$, or if the $\Delta \Sigma_l$ satisfy special properties.
But this is not generally the case.\footnote{
  Another disadvantage of Eq.~\eqref{eq:gobs_from_stacked} is that it cannot easily be generalized to stacking in $g_{\mathrm{bar}}$ space instead of position space.
  For the special case of baryonic point particles, this is simple enough; Eq.~\eqref{eq:gobs_from_esd} becomes $g_{\mathrm{obs}}(g_{\mathrm{bar}}) = 4G \int_0^{\pi/2} d\theta \Delta \Sigma(g_{\mathrm{bar}} \, \sin^2 \theta)$.
  For an individual galaxy, this can be generalized to other baryonic mass distributions as well.
  However, in general, the integral then depends on properties other than $g_{\mathrm{bar}}$ of the galaxy.
  This is a problem when applied to a stacked ESD profile $\Delta \Sigma^{\mathrm{stacked}}(g_{\mathrm{bar}})$ because then one no longer has access to properties of the individual galaxies other than $g_{\mathrm{bar}}$.
  In the following, we mostly consider baryonic point masses so this is only a secondary concern for us.
  We consider non-point masses only in Appendix~\ref{sec:appendix:SIShotgas}.
}

In practice, for the KiDS data we use (see Appendix~\ref{sec:appendix:newmethod:observational}), the weights $w_l(R)$ of the lenses $l$ all have roughly the same scaling with $R$.
Specifically, they all roughly scale as the projected area covered by the radial bin $R$.
Thus, the normalized weights $w_l(R) / \sum_{l'} w_{l'}(R)$ are roughly independent of $R$ (but they do depend on $l$) so that Eq.~\eqref{eq:gobs_from_stacked} works quite well.
Indeed, in practice, Eq.~\eqref{eq:gobs_from_stacked} gives results that are very close to what we get using the more general method we now propose, see Fig.~\ref{fig:rar-default-vs-from-stackedESD} in Appendix~\ref{sec:appendix:randomESD}.

Here, we want to use a stacked radial acceleration that \emph{is} an average of the radial accelerations of the individual lenses.
That is, we want our definition of $g_{\mathrm{obs}}^{\mathrm{stacked}}$ to be of the form Eq.~\eqref{eq:good_stacked_general}, i.e. $g_{\mathrm{obs}}^{\mathrm{stacked}}(R) = g_{\mathrm{obs}}^{\mathrm{averaged}}(R)$.
This can equivalently be written as
\begin{align}
 \label{eq:gobs_stacked_gobsl}
 g_{\mathrm{obs}}^{\mathrm{stacked}}(R) \equiv \bar{N}^{-1}(R) \sum_l \bar{w}_l(R) \, g_{\mathrm{obs},l}(R)\,,
\end{align}
where $g_{\mathrm{obs},l}(R)$ denotes the radial acceleration of each individual lens $l$.
The question then becomes how to choose the weights $\bar{w}_l(R)$.
Our choice is
\begin{align}
 \label{eq:gobs_weights}
 \bar{w}_l(R) \equiv \frac{1}{\left. \sigma_{g_{\mathrm{obs},l}}^2(R) \right|_{\mathrm{statistical}}}\,,
\end{align}
where $\left.\sigma_{g_{\mathrm{obs},l}}\right|_{\mathrm{statistical}}$ is the statistical uncertainty of the radial acceleration $g_{\mathrm{obs},l}$ of the lens $l$.
This choice is optimal in the sense that it minimizes the statistical uncertainty of the stacked radial acceleration $g_{\mathrm{obs}}^{\mathrm{stacked}}$, at least as long as the radial acceleration estimates $g_{\mathrm{obs},l}$ of the individual lenses can be considered independent.
We discuss this in more detail in Appendix~\ref{sec:appendix:errors} where we explain how we calculate statistical and systematic uncertainties.

Using  Eq.~\eqref{eq:gobs_stacked_gobsl} corresponds to first deprojecting individual galaxies and then stacking the result.
In contrast, Eq.~\eqref{eq:gobs_from_stacked} corresponds to first stacking the individual galaxies and then deprojecting.
As already mentioned, Eq.~\eqref{eq:gobs_from_stacked} has the disadvantage that it relies on the normalized weights being independent of $R$.
Since this is in general not the case, deprojecting after stacking is, in general, not exact.

But our method of choice -- to deproject first, then stack -- also has a disadvantage:
There is a restriction on how the various $\Delta \Sigma_l(R)$ values that one integrates over can be weighted relative to each other.
Indeed, for a given lens $l$ the relative weights of $\Delta \Sigma_l$ at different radii $R$ are fixed because one is not allowed to introduce an additional $R/\sin \theta$ dependence under the $\theta$ integral.
For example, consider one specific lens $l$.
If we know that $\Delta \Sigma_l(R_{\mathrm{good}})$ is much better measured than $\Delta \Sigma_l(R_{\mathrm{bad}})$, we would like to downweigh $\Delta \Sigma_l(R_{\mathrm{bad}})$ relative to $\Delta \Sigma_l(R_{\mathrm{good}})$.
But there is no such freedom here, since both occur under the same $\theta$ integral.
We can only adjust their common weight outside the integral.

Thus -- because bad data is not necessarily downweighted -- one may more easily be left with a systematic effect in the results with our method of choice.
Indeed, in Appendix~\ref{sec:appendix:randomESD}, we consider the signal from random coordinates and find a larger systematic trend using this method.
As usual in weak-lensing analyses, we deal with this systematic trend by subtraction \citep{Brouwer2021}, see Appendix~\ref{sec:appendix:errors}.
Still, one might contemplate simply using Eq.~\eqref{eq:gobs_from_stacked} instead because of this, i.e. stacking first and then deprojecting.
However, as discussed above, using Eq.~\eqref{eq:gobs_from_stacked} has its own systematic uncertainty from the fact that the normalized weights $N^{-1}(R) \, w_l(R)$ may depend on $R$.
In practice, both methods give very similar results, as we demonstrate in Appendix~\ref{sec:appendix:randomESD}.
We further validate both methods using the cross components of the ESD profiles in Appendix~\ref{sec:appendix:cross}.
Thus, we trust our results despite these systematic effects.

So far, we have considered stacking in position space.
That is, we have considered ESD profiles and radial accelerations that are averages over individual galaxies at a fixed radius $R$.
We are, however, also interested in things like the RAR that require the radial acceleration as a function of the Newtonian baryonic acceleration $g_{\mathrm{bar}}$,
\begin{align}
 g_{\mathrm{bar}}(R) \equiv \frac{G M_b(R)}{R^2} \,.
\end{align}
That is, we want to stack at fixed $g_{\mathrm{bar}}$ instead of at fixed $R$.
Note that each lens galaxy has a different baryonic mass $M_b(R)$, so we cannot easily get, for example, the ESD profile stacked in $g_{\mathrm{bar}}$ space, $\Delta \Sigma^{\mathrm{stacked}}(g_{\mathrm{bar}})$, from the ESD profile stacked in $R$ space, $\Delta \Sigma^{\mathrm{stacked}}(R)$.
Nevertheless, it is straightforward to adjust our stacking procedure to work in $g_{\mathrm{bar}}$ space, at least as long as there is a one-to-one mapping between radii $R$ and accelerations $g_{\mathrm{bar}}$ which we assume to be the case in the following.
We have for the stacked ESD profile
\begin{align}
 \label{eq:Rlofgbar}
 \Delta \Sigma^{\mathrm{stacked}}(g_{\mathrm{bar}}) = N^{-1}(g_{\mathrm{bar}}) \sum_l w_l(R_l(g_{\mathrm{bar}})) \, \Delta \Sigma_l(R_l(g_{\mathrm{bar}})) \,,
\end{align}
where $N(g_{\mathrm{bar}}) = \sum_l w_l(R_l(g_{\mathrm{bar}}))$ is a normalization factor and $R_l(g_{\mathrm{bar}})$ is the function that, for a given lens $l$, calculates the radius $R$ corresponding to a given Newtonian baryonic acceleration $g_{\mathrm{bar}}$.
If we approximate the baryonic mass as a point mass, this is
\begin{align}
 \label{eq:Rlofgbar_point}
 \left. R_l(g_{\mathrm{bar}}) \right|_{\mathrm{point}\;M_b} = \sqrt{\frac{G M_{b,l}}{g_{\mathrm{bar}}}} \,,
\end{align}
where $M_{b,l}$ is the baryonic mass of the lens $l$.
Other mass distributions require a different form of the function $R_l(g_{\mathrm{bar}})$, see for example Appendix~\ref{sec:appendix:SIShotgas}.
For the stacked radial acceleration, we similarly have,
\begin{align}
 \label{eq:gobs_stacked_ab}
 g_{\mathrm{obs}}^{\mathrm{stacked}}(g_{\mathrm{bar}}) \equiv 4G \bar{N}^{-1}(g_{\mathrm{bar}}) \sum_l \bar{w}_l(R_l(g_{\mathrm{bar}})) \int_0^{\pi/2} d \theta \Delta \Sigma_l \left(\frac{R_l(g_{\mathrm{bar}})}{\sin \theta}\right)\,.
\end{align}

\subsection{Stacking from observational data}
\label{sec:appendix:newmethod:observational}

We now explain how we obtain stacked ESD profiles and stacked radial accelerations from actual weak-lensing observations.
Essentially, we define the weights $w_l(R)$ and the ESD estimates $\Delta \Sigma_l(R)$ introduced in the previous section in terms of observational data.

We start with the stacked ESD profile written in the form given by Ref.~\cite{Brouwer2021},
\begin{align}
\label{eq:esd_stacked_Brouwer}
\Delta \Sigma^{\mathrm{stacked}}(R) = \frac{
  \sum_l \left. \sum_s\right|_{D_{ls} = R} W_{ls} \, \Sigma_{\mathrm{crit},ls} \epsilon_{t,ls}
}{
  \sum_l \left.\sum_s\right|_{D_{ls} = R} W_{ls}
} \,,
\end{align}
Here, the sum over the source galaxies $s$ runs over all sources within the radial bin $R$ of the lens $l$.
We indicate this in our notation by ``$D_{ls} = R$'' next to the summation sign.
The $W_{ls}$ are weights for each lens-source pair given by
\begin{align}
 W_{ls} = w_s \, \Sigma_{\mathrm{crit},ls}^{-2}  \,,
\end{align}
where $w_s$ estimates the precision of the ellipticity measurement of the source $s$ and $\Sigma_{\mathrm{crit},ls}$ is the critical surface density that we define below.
The tangential ellipticity $\epsilon_{t,ls}$ of the source $s$ is an estimate of the tangential shear $\gamma_t$ at a projected radius $R$ from the lens $l$ based on the second brightness moments of the source \citep{Bartelmann2001}.
It is given by
\begin{align}
\epsilon_{t,ls} = - \cos(2\phi_{ls}) \epsilon_{1,s} - \sin(2 \phi_{ls}) \epsilon_{2,s} \,,
\end{align}
where $\epsilon_{1,s}$ and $\epsilon_{2,s}$ are the ellipticity components of the source $s$ with respect to an equatorial coordinate system and $\phi_{ls}$ is the angle between the $x$-axis and the lens-source separation vector \citep{Dvornik2017,Bartelmann2001}.

The tangential shear is related to the ESD profile of the lens \citep{Bartelmann1995,Kaiser1995}.
In particular, the quantity $\Sigma_{\mathrm{crit},ls} \epsilon_{t,ls}$ is an estimate for the ESD of the lens $l$ at radius $R$ based on the tangential ellipticity $\epsilon_{t,ls}$ of the source $s$,
\begin{align}
 \left. \Delta \Sigma_l\right|_{\mathrm{from}\;s}(R = D_{ls}) = \Sigma_{\mathrm{crit},ls} \, \epsilon_{t,ls} \,.
\end{align}
By taking a weighted average over the sources $s$, we get an estimate for the ESD of the lens $l$ in a radial bin $R$ that includes all relevant sources,
\begin{align}
 \label{eq:esd_l}
 \Delta \Sigma_l(R) = \frac{\left.\sum_s\right|_{D_{ls} = R} W_{ls} \, \Sigma_{\mathrm{crit},ls} \, \epsilon_{t,ls}}{\left.\sum_s\right|_{D_{ls} = R} W_{ls}} \,.
\end{align}
If we now define
\begin{align}
 \label{eq:w_l}
 w_l(R) = \left.\sum_s\right|_{D_{ls} = R} W_{ls} \,,
\end{align}
we can write the original formula Eq.~\eqref{eq:esd_stacked_Brouwer} for the stacked ESD in the following form,
\begin{align}
 \Delta \Sigma^{\mathrm{stacked}}(R) = N^{-1}(R) \sum_l w_l(R) \, \Delta \Sigma_l(R) \,,
\end{align}
with the normalization factor $N(R) = \sum_l w_l(R)$.
This is the form we already used in the previous section in Eq.~\eqref{eq:esd_stacked}.
Thus, Eq.~\eqref{eq:esd_l} and Eq.~\eqref{eq:w_l} are the definitions of $\Delta \Sigma_l$ and $w_l$ in terms of observational quantities.

It remains to give the definition of the critical surface density $\Sigma_{\mathrm{crit},ls}$.
If one knows the exact redshifts of both the source, $z_s$, and the lens, $z_l$, one has
\begin{align}
 \Sigma^{-1}_{\mathrm{crit},ls} = \frac{4 \pi G}{c^2} \frac{D(z_l) D(z_l, z_s)}{D(z_s)} \,.
\end{align}
In practice, however, we use a more complicated definition of $\Sigma_{\mathrm{crit},ls}$ that takes observational uncertainties into account.
We discuss this in Sec.~\ref{sec:data}; see Eq.~\eqref{eq:sigmacrit_full}.

For the continuous integrals in the deprojection formula Eq.~\eqref{eq:gobs_from_esd} (see also Eq.~\eqref{eq:good_stacked_general}), we need to know $\Delta \Sigma_l(R)$ for all values of $R$, not just for discrete radial bins, which is what we have discussed so far.
For simplicity, we linearly interpolate $\Delta \Sigma_l$ between these discrete bins (see Appendix~\ref{sec:appendix:errors} for how we estimate the uncertainty associated with this interpolation).
When there are no sources in some radial bin, we do not have an estimate of $\Delta \Sigma_l$ there.
In such cases, we linearly interpolate between the bins that do have sources.

\section{Uncertainties and cross-checks}

\subsection{Statistical and systematic uncertainties}
\label{sec:appendix:errors}

Here, we discuss the statistical and systematic uncertainties of our stacked radial accelerations $\gobs$ and stacked ESD profiles $\Delta \Sigma$.
In the following, we show explicit expressions only for quantities stacked in position space.
We use the same procedure when stacking in $g_{\mathrm{bar}}$ space.
We first repeat the formula Eq.~\eqref{eq:esd_stacked} for our stacked ESD profiles $\Delta \Sigma(R)$ to make it easier for the reader to follow the rest of this section.
We have
\begin{align}
 \Delta \Sigma^{\mathrm{stacked}}(R) = N^{-1}(R) \sum_l w_l(R) \, \Delta \Sigma_l(R) \,,
\end{align}
where $N(R) = \sum_l w_l(R)$ normalizes the weights $w_l(R)$.
The weights $w_l(R)$ and the ESD profiles $\Delta \Sigma_l(R)$ of the lens $l$ are defined in terms of observational data in Eq.~\eqref{eq:esd_l} and Eq.~\eqref{eq:w_l}.
We similarly repeat the formula Eq.~\eqref{eq:gobs_stacked_gobsl} for our stacked radial accelerations $\gobs(R)$.
We have
\begin{align}
 g_{\mathrm{obs}}^{\mathrm{stacked}}(R) = \bar{N}^{-1}(R) \sum_l \bar{w}_l(R) \, g_{\mathrm{obs},l}(R)\,,
\end{align}
where $\bar{N}(R) = \sum_l \bar{w}_l(R)$ normalizes the weights $\bar{w}_l(R)$ that are defined as the inverse of the squared statistical uncertainty of $g_{\mathrm{obs},l}$, i.e. $\bar{w}_l(R) = 1/\sigma^2_{g_{\mathrm{obs}},l}(R)$.
We will define this statistical uncertainty in Eq.~\eqref{eq:gobsl_stacked_err_stat} below.
The radial acceleration $g_{\mathrm{obs},l}$ of the lens $l$ is given by
\begin{align}
 \label{eq:gobsldefinition}
 g_{\mathrm{obs},l}(R) = 4G \int_0^{\pi/2} d\theta \Delta \Sigma_l\left(\frac{R}{\sin \theta}\right) \,.
\end{align}

We now consider the statistical uncertainties of our stacked ESD profiles and stacked radial accelerations.
We start with the ESD profile estimates $\Delta \Sigma_l$ for individual lenses $l$ from Eq.~\eqref{eq:esd_l}.
For these, we use the analytical error estimate used by Refs.~\cite{Brouwer2021,Viola2015}, but adjust it for the fact that we are dealing with only a single lens for now.
We find (see Appendix~\ref{sec:appendix:newmethod:observational} for the definition of $W_{\mathrm{ls}}$ and $\Sigma_{\mathrm{crit},ls}$),
\begin{align}
 \label{eq:sigma_esd_l}
 \left. \sigma_{\Delta \Sigma_l}^2(R)\right|_{\mathrm{statistical}} = \frac{\left.\sum_s \right|_{D_{ls} = R} W_{ls}^2 \, \Sigma_{\mathrm{crit},ls}^2 \, \sigma_{\epsilon,s}^2}{\left( \left. \sum_s \right|_{D_{ls} = R} W_{ls} \right)^2} \,,
\end{align}
where we take the ellipticity dispersion $\sigma_{\epsilon,s}$ from Table I in Ref.~\cite{Giblin2021} which lists values for five tomographic redshift bins.
We choose the one that corresponds to the redshift of the source $s$.
This expression follows directly from Eq.~\eqref{eq:esd_l} and is simpler than the one derived in Ref.~\cite{Viola2015}, even when considering only the diagonal elements of their covariance matrix.
This is because each source occurs at most once in Eq.~\eqref{eq:esd_l} since we consider only a single lens for now.

To obtain statistical uncertainties for the stacked ESD profiles and the stacked radial accelerations, we linearly propagate these $\sigma_{\Delta \Sigma_l}$ uncertainties, assuming that the $\Delta \Sigma_l(R)$ are independent for different lenses $l$ and different radii $R$.
In particular, this assumes that only very few sources contribute to multiple lenses.
In contrast, Refs.~\cite{Brouwer2021,Viola2015} take such cases into account.
However, since our analysis -- like that of Ref.~\cite{Brouwer2021} -- relies on lenses being isolated, such cases should not be important.
Indeed, we have verified that our simplified procedure reproduces the statistical uncertainties of Ref.~\cite{Brouwer2021} almost perfectly.

Concretely, we adopt the following statistical uncertainties for our stacked ESD profiles from Eq.~\eqref{eq:esd_stacked},
\begin{align}
 \left. \sigma_{\Delta \Sigma^{\mathrm{stacked}}}^2(R)\right|_{\mathrm{statistical}} = N^{-2}(R) \sum_l w_l^2(R) \, \left. \sigma^2_{\Delta \Sigma_l}(R)\right|_{\mathrm{statistical}}\,.
\end{align}
And similarly for our stacked radial accelerations from Eq.~\eqref{eq:gobs_stacked_gobsl},
\begin{align}
 \label{eq:gobs_stacked_err_stat}
 \left. \sigma_{g^{\mathrm{stacked}}_{\mathrm{obs}}}^2(R)\right|_{\mathrm{statistical}} = \bar{N}^{-2}(R) \sum_l \bar{w}_l^2(R) \left. \sigma_{g_{\mathrm{obs},l}}^2(R)\right|_{\mathrm{statistical}} \,,
\end{align}
with
\begin{align}
 \label{eq:gobsl_stacked_err_stat}
 \left. \sigma_{g_{\mathrm{obs},l}}^2(R)\right|_{\mathrm{statistical}} =  (4G)^2 \int_0^{\pi/2} d\theta \left. \sigma^2_{\Delta \Sigma_l}\left(\frac{R}{\sin \theta}\right)\right|_{\mathrm{statistical}}\,.
\end{align}
The $\theta$ integral involves $\sigma_{\mathrm{\Delta \Sigma_l}}$ at arbitrarily large radii and, as discussed in Sec.~\ref{sec:newmethod:individual}, we set $\sigma_{\Delta \Sigma_l}$ to zero beyond the last $\Delta \Sigma_l$ data point.

In practice, we know $\Delta \Sigma_l(R)$ only in a finite number of discrete radial bins, see Appendix~\ref{sec:appendix:newmethod:observational}.
As discussed in Sec.~\ref{sec:newmethod:individual}, this means there are systematic uncertainties related to both extrapolating beyond the last bin and interpolating between the discrete bins.
For our stacked radial accelerations, we estimate these systematic uncertainties in the same way as for individual galaxies, see Sec.~\ref{sec:newmethod:individual}.
That is, we adopt the systematic uncertainties $\sigma_{g_{\mathrm{obs}}^{\mathrm{stacked}}}|_{\mathrm{systematic}}$ from Eq.~\eqref{eq:syst_err_gobs_extrapolate} and Eq.~\eqref{eq:syst_err_gobs_interpolate}.

Following Ref.~\cite{Brouwer2021}, we correct for both additive and multiplicative biases.
In Ref.~\cite{Brouwer2021}, it was found that the multiplicative bias is independent of the projected distance from the lens.
Thus, we can correct our stacked ESD profiles, stacked radial accelerations, and their uncertainties at the very end of our calculation,
\begin{align}
 \label{eq:bias_mult}
 \Delta \Sigma^{\mathrm{stacked}} \to \frac{1}{1+ \mu} \Delta \Sigma^{\mathrm{stacked}} \,, \quad
 g_{\mathrm{obs}}^{\mathrm{stacked}} \to \frac{1}{1+ \mu}g_{\mathrm{obs}}^{\mathrm{stacked}} \,, \quad
\end{align}
where we adopt the value $1+\mu = 0.98531$ provided by Ref.~\cite{Brouwer2021} in their data files.
For brevity, Eq.~\eqref{eq:bias_mult} shows the multiplicative correction only for the ESD profiles and radial accelerations, but the same procedure applies also to their respective uncertainties.

For the additive biases, we follow Ref.~\cite{Brouwer2021} and calculate the ESD profile of a set of random coordinates with random redshifts.
When calculating a stacked ESD profile, we then subtract the ``random'' ESD profile $\Delta \Sigma^{\mathrm{random}}(R)$ from the $\Delta \Sigma_l(R)$ of each lens $l$,
\begin{align}
 \Delta \Sigma_l(R) \to \Delta \Sigma_l(R)  - \Delta \Sigma^{\mathrm{random}}(R) \,.
\end{align}
When calculating stacked radial accelerations, we similarly subtract the ``random'' radial acceleration $g_{\mathrm{obs}}^{\mathrm{random}}(R)$ from the $g_{\mathrm{obs},l}(R)$ of each lens $l$,\footnote{
 Since $g_{\mathrm{obs},l}$ is ultimately calculated from an integral over $\Delta \Sigma_l$, see Eq.~\eqref{eq:gobsldefinition}, one might instead consider just subtracting $\Delta \Sigma^{\mathrm{random}}$ from $\Delta \Sigma_l$ inside that integral.
 However, this essentially means switching from ``deproject first, then stack'' to ``stack first, then deproject'' for the subtracted random profile.
 As explained in Appendix~\ref{sec:appendix:newmethod:stacked}, this difference can be important for systematic effects such as those that we aim to subtract here.
 Thus, Eq.~\eqref{eq:gobslsubtract} is the preferred way of doing this subtraction.
}
\begin{align}
 \label{eq:gobslsubtract}
 g_{\mathrm{obs},l}(R) \to g_{\mathrm{obs},l}(R)  - g_{\mathrm{obs}}^{\mathrm{random}}(R) \,.
\end{align}
How we calculate $\Delta \Sigma^{\mathrm{random}}$ and $g_{\mathrm{obs}}^{\mathrm{random}}$ is explained in Appendix~\ref{sec:appendix:randomESD}.

\subsection{Additive bias from random coordinates}
\label{sec:appendix:randomESD}

Here, we describe how we calculate the additive bias that we subtract when calculating our stacked ESD profiles and stacked radial accelerations (see Appendix~\ref{sec:appendix:errors}).
We mostly follow the procedure of Ref.~\cite{Brouwer2021}, with some adjustments for our different method of deriving radial accelerations from weak lensing data.

We first generate about 45 million uniform random coordinates -- exactly 50 times the size of the KiDS-bright sample when not applying any mass cut or isolation criterion -- with the footprint of the 1006 KiDS DR4 tiles considered by Ref.~\cite{Brouwer2021}.
Then, we divide the KiDS-bright sample into 80 linear redshift bins between $0.1$ and $0.5$.
For each of these bins, we multiply the number of galaxies in that bin by 50 and generate that many random uniform redshifts in that bin.
Finally, we assign the random redshifts to the random coordinates and measure the stacked ESD profile $\Delta \Sigma^{\mathrm{stacked}}$, the stacked ESD cross profile $\Delta \Sigma^{\mathrm{stacked}}_\times$, the stacked radial acceleration $g_{\mathrm{obs}}^{\mathrm{stacked}}$, and the stacked radial acceleration cross profile $g_{\mathrm{obs},\times}^{\mathrm{stacked}}$.
For this, we use the same procedure we use for the actual lenses (see Appendix~\ref{sec:appendix:newmethod:observational}).
We use 27 logarithmic radial bins between $0.003\,\mathrm{Mpc}/h_{70}$ and $11.94\,\mathrm{Mpc}/h_{70}$.
The cross profiles were not discussed in Appendix~\ref{sec:appendix:newmethod:observational} because they were not needed there.
We treat the cross profiles in the same way as the tangential ones with two changes.
First, the cross profile should ideally be zero.
Thus, we assume that $\Delta \Sigma_{l,\times}$ is zero beyond the last data point rather than continuing with an SIS profile.
Second, the cross profiles are obtained from the cross components $\epsilon_{\times,ls}$ of the ellipticities instead of the  tangential components $\epsilon_{t,ls}$.
Using the notation of Appendix~\ref{sec:appendix:newmethod:observational}, we have
\begin{align}
 \epsilon_{\times,ls} = \sin(2\phi_{ls}) \epsilon_{1,s} - \cos(2 \phi_{ls}) \epsilon_{2,s} \,.
\end{align}

\begin{figure}
 \centering
  \includegraphics[width=.49\textwidth]{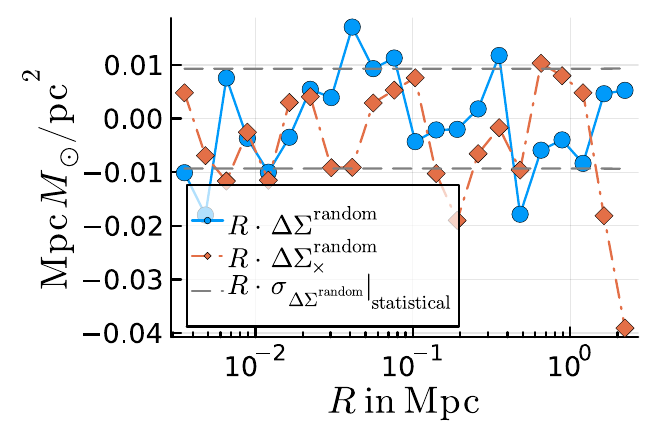}
  \includegraphics[width=.49\textwidth]{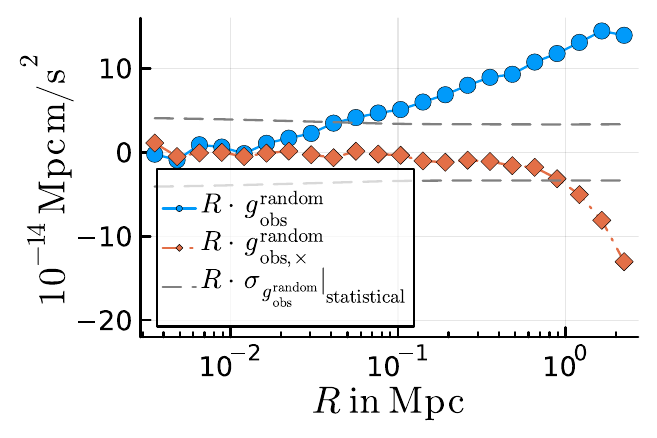}
 \caption{
   The tangential and cross components of the ESD profile (left) and the radial acceleration (right) for about 45 million random coordinates for $R < 3\,\mathrm{Mpc}$ (see Fig.~\ref{fig:random-ESD-inf} for larger radii), assuming $h_{70} = 1$.
   These should ideally be close to zero.
   As discussed in the text, the statistical uncertainties shown here are only a rough estimate and should not be taken too seriously.
   That the tangential radial acceleration starts to systematically deviate from zero already at relatively small radii is likely due to the effect discussed in Appendix~\ref{sec:appendix:newmethod:stacked}, namely that one cannot downweigh bad data as optimally as for the ESD profile when using our preferred method (i.e. deproject first, then stack).
   This is supported by the fact that using Eq.~\eqref{eq:gobs_from_esd} to convert the tangential ESD profile from the left panel to tangential radial accelerations (i.e. stack first, then deproject) does not reproduce the systematic trend in the right panel.
   We nevertheless trust our results because, after subtraction, the radial accelerations calculated using our ``deproject first, then stack" method agree very well with those obtained using the ``stack first, then deproject'' method which has very different systematics (see Fig.~\ref{fig:rar-default-vs-from-stackedESD}).
 }
 \label{fig:random-ESD-3}
\end{figure}

\begin{figure}
 \centering
  \includegraphics[width=.49\textwidth]{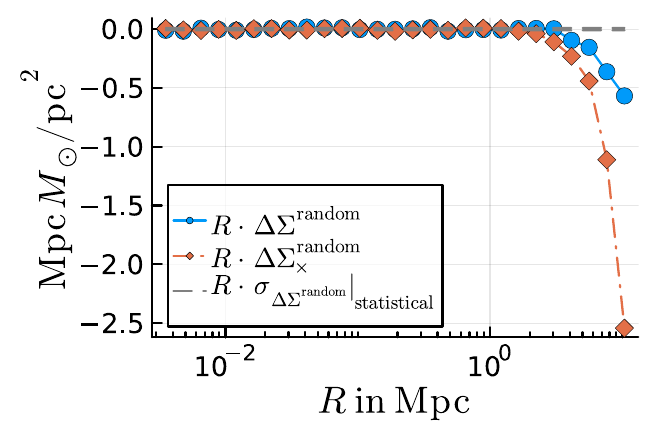}
  \includegraphics[width=.49\textwidth]{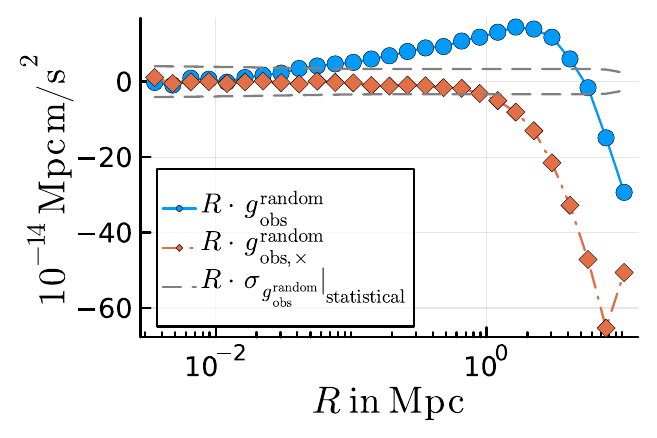}
 \caption{
   Same as Fig.~\ref{fig:random-ESD-3} but not restricted to $R < 3\,\mathrm{Mpc}$.
 }
 \label{fig:random-ESD-inf}
\end{figure}

Both tangential and cross profiles should ideally be close to zero for random coordinates.
From Fig.~\ref{fig:random-ESD-3} and Fig.~\ref{fig:random-ESD-inf}, we see that there are systematic non-zero residuals at large radii.
Such systematic effects at large radii are not uncommon, see for example Appendix~A of Ref.~\cite{Dvornik2017}.
This effect is strongest for the tangential radial acceleration, for which the systematic deviation from zero starts at smaller radii than for the other profiles shown.
As discussed in Appendix~\ref{sec:appendix:newmethod:stacked}, the radial acceleration is particularly prone to such effects because one cannot downweigh unreliable data as well as for the ESD profiles.
We suggest that this effect explains why the tangential radial acceleration shows a systematic deviation from zero already at relatively small radii.

This interpretation is supported by the following observation.
Our radial accelerations are based on the integral formula Eq.~\eqref{eq:gobs_from_esd}.
We have verified that applying this formula to the stacked ESD profile obtained from random coordinates (Fig.~\ref{fig:random-ESD-3}, left) does \emph{not} reproduce the systematic trend we see in the tangential radial acceleration (Fig.~\ref{fig:random-ESD-3}, right).
Thus, neither the data nor the integral formula by themselves are responsible for this systematic trend.
But it may well be that the restrictions on downweighing bad data discussed in Appendix~\ref{sec:appendix:newmethod:stacked} are.

\begin{figure}
 \centering
  \includegraphics[width=.9\textwidth]{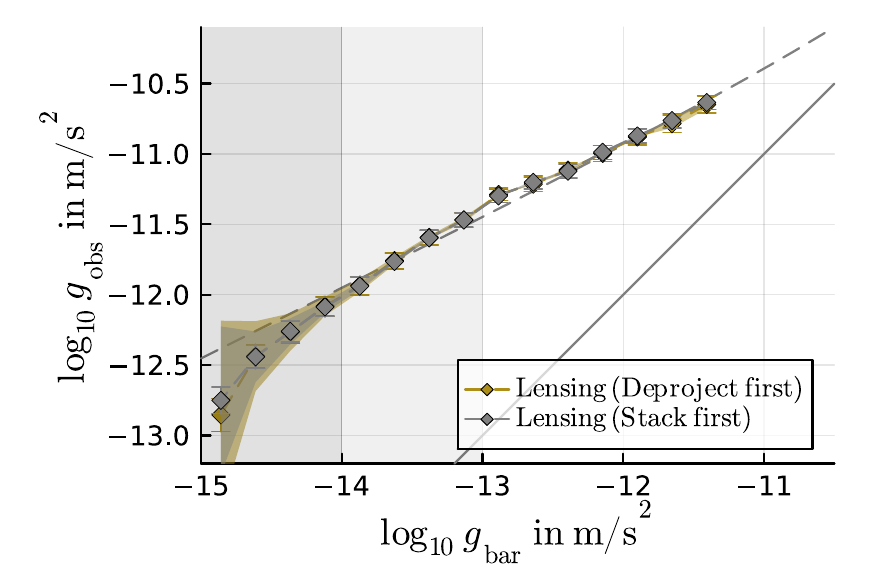}
 \caption{
   The weak-lensing RAR obtained using our method of choice as described in Appendix~\ref{sec:appendix:newmethod} (i.e. deproject first, then stack, yellow diamonds) and using Eq.~\eqref{eq:gobs_from_stacked} (i.e. stack first, then deproject, gray diamonds).
   Since Eq.~\eqref{eq:gobs_from_stacked} uses the stacked ESD profile as its input, it is not affected by the systematic trend for the tangential radial acceleration of random coordinates shown in Fig.~\ref{fig:random-ESD-3}, right, but comes with its own systematic uncertainties instead (see Appendix~\ref{sec:appendix:newmethod:stacked}).
   That these two procedures with different systematics agree so well indicates that our results can be trusted.
   Error bars and bands are as in Fig.~\ref{fig:rar}, except we do not show the stellar mass systematic uncertainty for clarity.
 }
 \label{fig:rar-default-vs-from-stackedESD}
\end{figure}

We trust our results despite this systematic trend because -- after subtracting the profile obtained from the random coordinates -- the $\gobs$ obtained using our method of choice (i.e. deproject first, then stack) agrees very well with that obtained using Eq.~\eqref{eq:gobs_from_stacked}\footnote{
  The expression Eq.~\eqref{eq:gobs_from_stacked} applies when stacking in position space.
  As mentioned in Appendix~\ref{sec:appendix:newmethod:stacked}, this is in general not straightforward to adapt to stacking in $g_{\mathrm{bar}}$ space.
  However, as also mentioned there, this \emph{is} straightforward for baryonic point particles, which is what we are interested here.
  Thus, for Fig.~\ref{fig:rar-default-vs-from-stackedESD} we adapt Eq.~\eqref{eq:gobs_from_stacked} in the following way, $g_{\mathrm{obs}}(g_{\mathrm{bar}}) = 4G \int_0^{\pi/2} d\theta \, \Delta \Sigma^{\mathrm{stacked}}(g_{\mathrm{bar}} \sin^2 \theta)$.
} (i.e. stack first, then deproject), see Fig.~\ref{fig:rar-default-vs-from-stackedESD}.
Importantly, as explained in Appendix~\ref{sec:appendix:newmethod}, the ``stack first, then deproject'' method from Eq.~\eqref{eq:gobs_from_stacked} is not prone to the systematic effect we consider here because it uses the stacked and subtracted ESD profile as input which does not show such a systematic trend at relatively small radii, see Fig.~\ref{fig:random-ESD-3}.
Fig.~\ref{fig:rar-default-vs-from-stackedESD} shows some difference at the very last data point.
But this is where systematic uncertainties are anyway large, so in practice the difference in this last data point is not important.

The statistical uncertainties shown in Fig.~\ref{fig:random-ESD-3} and Fig.~\ref{fig:random-ESD-inf} are calculated as described in Appendix~\ref{sec:appendix:errors}.
In particular, the calculation assumes that sources do not contribute to multiple lenses.
That is a good approximation for the isolated lenses we consider in the main text.
But it is not justified for the large sample of random coordinates we consider here.
Thus, the numerical values of the statistical uncertainties shown in Fig.~\ref{fig:random-ESD-3} and Fig.~\ref{fig:random-ESD-inf} give only a rough indication of the order of magnitude of the uncertainties.
Note that these uncertainties do not enter any of our results in the main text.
They are used only to guide the eye of the reader in Fig.~\ref{fig:random-ESD-3} and Fig.~\ref{fig:random-ESD-inf}.

\subsection{Cross component from actual lenses}
\label{sec:appendix:cross}

As discussed in Appendix~\ref{sec:appendix:randomESD}, both the cross and tangential components of the ESD profiles and radial accelerations should be zero for random coordinates.
For actual lenses, the tangential component carries the lensing signal and should not be zero, while the cross component should still be zero.
This can be used to validate the lensing data and the method used to analyse this data.
In the following, we subtract the cross components obtained from random coordinates (see Appendix~\ref{sec:appendix:randomESD}) from the cross components of the stacked ESD profiles and stacked radial accelerations, just as we do for the tangential components (see Appendix~\ref{sec:appendix:errors}).

\begin{figure}
 \centering
  \includegraphics[width=\textwidth]{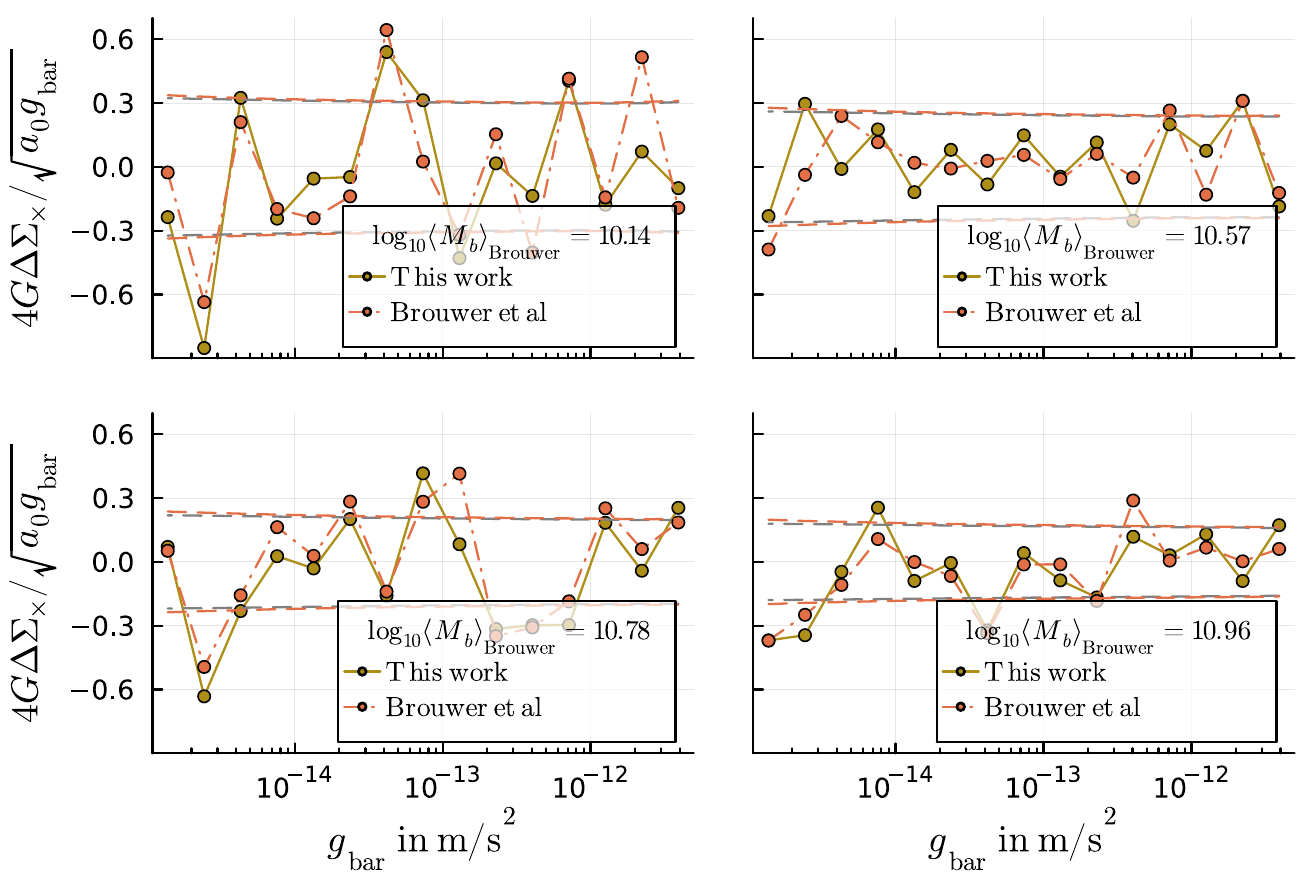}
  \caption{
    The cross components of the stacked ESD profiles for the four stellar mass bins defined by Ref.~\cite{Brouwer2021} as derived in this work (yellow) and in Ref.~\cite{Brouwer2021} (red).
    To allow a direct comparison to Ref.~\cite{Brouwer2021}, for this plot we adopt their $h_{70} = 1$, their stellar and gas masses, and their cuts on the KiDS bright sample, namely $R_{\mathrm{isol}} = 3\,\mathrm{Mpc}/h_{70}$ and $\log_{10} M_\ast/M_\odot < 11$.
  }
  \label{fig:cross-ESD-Brouwer}
\end{figure}

We first consider the cross-components of the stacked ESD profiles.
Since the version of these cross ESD profiles from Ref.~\cite{Brouwer2021} is publicly available, we can directly compare our results to theirs.
For this, we adopt the stellar and gas masses, stellar mass bins, choice of $h_{70}$, and cuts on the KiDS bright sample from Ref.~\cite{Brouwer2021}.
We show the result in Fig.~\ref{fig:cross-ESD-Brouwer}.
We see that our results match those of Ref.~\cite{Brouwer2021} very closely.
There are some differences, but these are much smaller than the error bars and may be due to any number of minor numerical differences.
We also see that these cross ESD profiles are consistent with zero, as they should be.

\begin{figure}
 \centering
  \includegraphics[width=\textwidth]{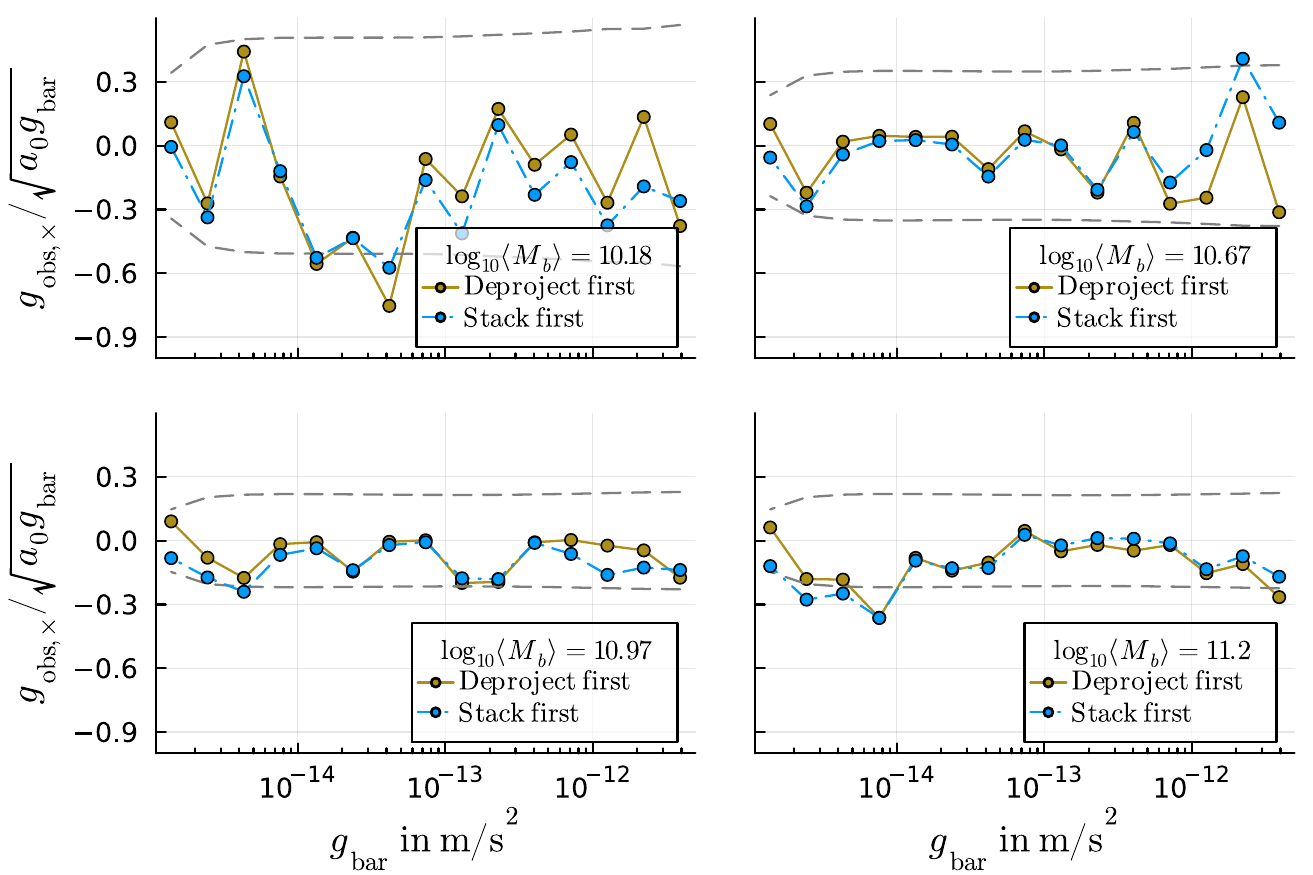}
  \caption{
    The cross components of the stacked radial accelerations for the four baryonic mass bins defined in Sec.~\ref{sec:massbins} derived using our method described in Sec.~\ref{sec:newmethod} (i.e. deproject first, then stack, yellow).
    For comparison, we also show the cross components of the radial accelerations derived from the stacked ESD profile and Eq.~\eqref{eq:gobs_from_stacked} (i.e. stack first, then deproject, blue, see also Fig.~\ref{fig:rar-default-vs-from-stackedESD}).
  }
  \label{fig:cross-gobs}
\end{figure}

In Fig.~\ref{fig:cross-gobs} we show the cross components of the stacked radial accelerations obtained using our method for converting ESD profiles to accelerations described in Sec.~\ref{sec:newmethod} (i.e. deproject first, then stack).
In contrast to Fig.~\ref{fig:cross-ESD-Brouwer}, we here use our own masses, mass bins, choice of $h_{70}$, and cuts on the KiDS bright sample (see Sec.~\ref{sec:masses}, Sec.~\ref{sec:data} and Sec.~\ref{sec:massbins}).
For comparison, we also show the cross components of the radial accelerations obtained using Eq.~\eqref{eq:gobs_from_stacked} (i.e. stack first, then deproject).
We see that both methods agree well with each other (see also Fig.~\ref{fig:rar-default-vs-from-stackedESD}) and produce a cross component consistent with zero.

There is a slight tendency for the cross components in Fig.~\ref{fig:cross-gobs} to fall below zero more often than above zero.
However, this may well just be random fluctuations.
Indeed, the cross components do generally stay within the error bars.
In addition, we have verified that if we derive the cross-component of random coordinates as in Appendix~\ref{sec:appendix:randomESD} but with a much smaller number of random coordinates matching our sample of actual lenses, it is not hard to find similar behavior, even after subtracting the cross components inferred from the (much larger) full random sample from Appendix~\ref{sec:appendix:randomESD}.

\subsection{Influence of larger radii on smaller radii}
\label{sec:appendix:cutab14}

\begin{figure}
 \centering
  \includegraphics[width=.9\textwidth]{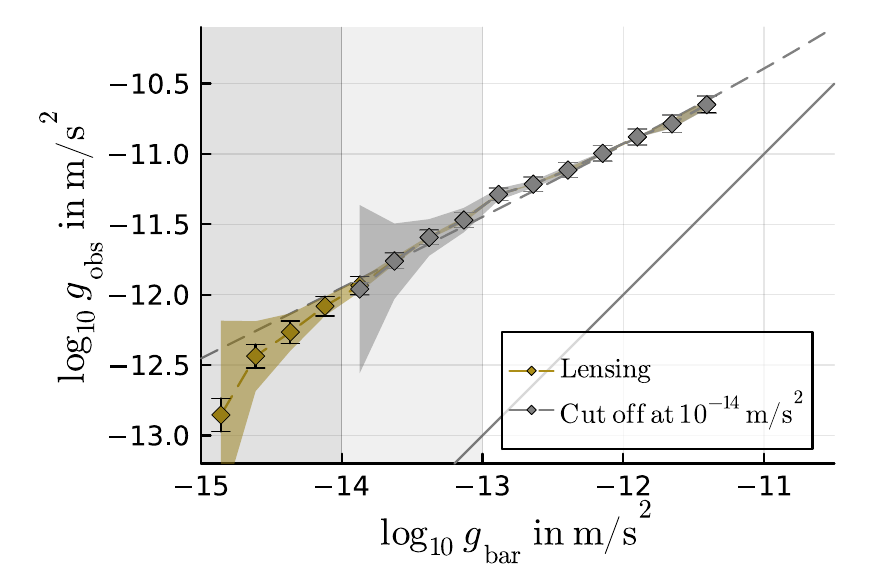}
  \caption{
    The weak-lensing RAR derived using the deprojection formula Eq.~\eqref{eq:gobs_from_esd} when using the full ESD profiles of the individual lenses (as in Fig.~\ref{fig:rar}, yellow diamonds) and when artificially cutting off the data at $g_{\mathrm{bar}} = 10^{-14}\,\mathrm{m/s}^2$ (gray diamonds).
    We find very similar results in both cases.
    Thus, if the isolation criterion fails below $g_{\mathrm{bar}} = 10^{-14}\,\mathrm{m/s}^2$, this has only very little effect on $g_{\mathrm{obs}}$ at larger values of $g_{\mathrm{bar}}$.
    Error bars and bands are as in Fig.~\ref{fig:rar}, except we do not show the stellar mass systematic uncertainty for clarity.
  }
 \label{fig:rar-cutab14}
\end{figure}

As discussed in Sec.~\ref{sec:rar}, the deprojection formula Eq.~\eqref{eq:gobs_from_esd} involves data at arbitrarily large radii.
Thus, one may worry that, if the isolation criterion fails at large radii, radial accelerations at all radii will be affected, not just those at large radii.
Indeed, our results from Sec.~\ref{sec:etgltg} suggest that our isolation criterion is probably not reliable below $\gbar \approx 10^{-14}\,\mathrm{m/s}^2$.
Fig.~\ref{fig:rar-cutab14} shows that this is not a problem in practice.
We find very similar results when we artificially cut off the ESD profiles $\Delta \Sigma_l$ of the individual lenses at $g_{\mathrm{bar}} = 10^{-14}\,\mathrm{m/s}^2$ instead of using data down to $g_{\mathrm{bar}} = 10^{-15}\,\mathrm{m/s}^2$.
We have verified that the same is true if we artificially cut off the data at $\gbar = 10^{-13}\,\mathrm{m/s}^2$.
Mathematically, this is because most of the integration volume of the integral in Eq.~\eqref{eq:gobs_from_esd} is close to $R$, i.e. large radii are downweighted relative to radii close to $R$.

\subsection{$R_{\mathrm{isol}}$ dependence with artificially fixed error bars}
\label{sec:appendix:rescalederrors}

\begin{figure}
 \centering
  \includegraphics[width=.8\textwidth]{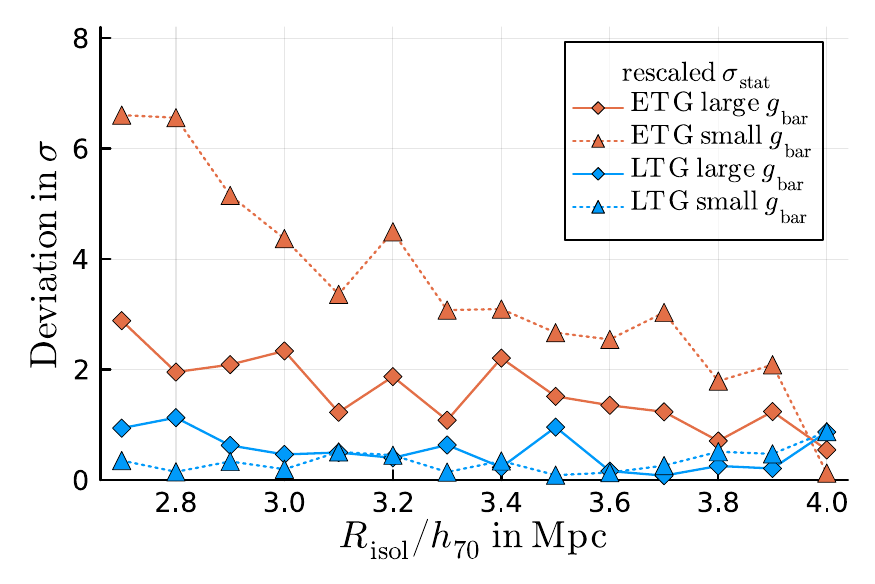}
 \caption{
   Same as the top panel of Fig.~\ref{fig:etgltg-sigmas}, but with artificially rescaled error bars.
   The idea is that, at larger $R_{\mathrm{isol}}$, the lens sample is smaller so the error bars are larger.
   Thus, the number of sigmas at larger $R_{\mathrm{isol}}$ may be small simply because of the larger error bars.
   To counter this, we artificially rescale the error bars to the level they were at for $R_{\mathrm{isol}} = 3.0\,\mathrm{Mpc}/h_{70}$, see Eq.~\eqref{eq:sigma_sqrtN_rescale}.
   Even with these artificially small error bars, the qualitative trend with $R_{\mathrm{isol}}$ remains the same as in Fig.~\ref{fig:etgltg-sigmas}.
 }
 \label{fig:etgltg-sigmas-rescaled-errors}
\end{figure}

In Sec.~\ref{sec:etgltg}, we saw that ETGs are quite sensitive to the details of the isolation criterion as quantified by $R_{\mathrm{isol}}$, while, at least down to $\gbar \approx 10^{-14}\,\mathrm{m/s}^2$, LTGs are almost independent of $R_{\mathrm{isol}}$.
In particular, we considered $R_{\mathrm{isol}}$ up to $4\,\mathrm{Mpc}/h_{70}$, which gives a stricter isolation criterion than that used in Ref.~\cite{Brouwer2021}, namely $R_{\mathrm{isol}} = 3\,\mathrm{Mpc}/h_{70}$.
For large values of $R_{\mathrm{isol}}$, we found no significant difference between ETGs and LTGs.

A stricter isolation criterion implies a smaller lens sample and thus larger statistical uncertainties.
Thus, in principle, the reason we find no difference between ETGs and LTGs may simply be that it becomes statistically insignificant because of the larger uncertainties.
To counter this, Fig.~\ref{fig:etgltg-sigmas-rescaled-errors} shows what happens when we artificially keep the error bars fixed at the level they are at for $R_{\mathrm{isol}} = 3\,\mathrm{Mpc}/h_{70}$.

More specifically, we have verified that the statistical uncertainties scale  as $1/\sqrt{N}$ to an excellent approximation.
Here, $N$ is the number of lenses in  the sample.
Thus, to keep the error bars at the level they are at for $R_{\mathrm{isol}} = 3\,\mathrm{Mpc}/h_{70}$, we simply adjust the statistical uncertainties in the following way before calculating $\chi^2$,
\begin{align}
 \label{eq:sigma_sqrtN_rescale}
 \left.\sigma_{g_{\mathrm{obs}}}\right|_{\mathrm{statistical}} \to
 \left.\sigma_{g_{\mathrm{obs}}}\right|_{\mathrm{statistical}} \times \sqrt{\frac{N}{N_{3.0}}} \,,
\end{align}
where $N$ is the number of lenses that satisfy the isolation criterion with the value of $R_{\mathrm{isol}}$ that is currently under consideration and $N_{3.0}$ is the number of lenses that satisfy the isolation criterion with $R_{\mathrm{isol}} = 3\,\mathrm{Mpc}/h_{70}$.
Comparing Fig.~\ref{fig:etgltg-sigmas} and Fig.~\ref{fig:etgltg-sigmas-rescaled-errors} shows that our results remain unchanged even with these artificially small statistical uncertainties.

\subsection{Extended hot gas distribution}
\label{sec:appendix:SIShotgas}

As discussed in Sec.~\ref{sec:masses}, we assume that ETGs are surrounded by hot gas and we model this hot gas as a point mass.
In reality, however, this hot gas may be quite extended.
To illustrate the effect of this, here we follow Ref.~\cite{Brouwer2021} and model the hot gas around ETGs as having an SIS profile cut off at $100\,\mathrm{kpc}$.

To obtain radial accelerations stacked in $g_{\mathrm{bar}}$ space using this extended mass profile, we must modify the functions $R_l(g_{\mathrm{bar}})$ that map between radii $R_l$ and baryonic Newtonian accelerations $g_{\mathrm{bar}}$ for each lens $l$, see Eq.~\eqref{eq:Rlofgbar} and Eq.~\eqref{eq:Rlofgbar_point}.
In particular, the point mass relation Eq.~\eqref{eq:Rlofgbar_point} remains valid for LTGs, but for ETGs we now have
\begin{align}
\left. R_l(g_{\mathrm{bar}}) \right|_{\mathrm{ETG},\mathrm{SIS}} =
\begin{cases}
 \sqrt{
   \frac{G (M_{\ast,l} + M_{\mathrm{gas},l})}{g_{\mathrm{bar}}}
  }, & \mathrm{for}\; g_{\mathrm{bar}} < a_{c,l} \,, \\
 R_c \, \frac{a_{g,l}}{2 g_{\mathrm{bar}}} \left(
  1 + \sqrt{1 + 4 \frac{g_{\mathrm{bar}}}{a_{g,l}} \frac{M_{\ast,l}}{M_{\mathrm{gas},l}} }
  \right), & \mathrm{for}\; g_{\mathrm{bar}} \geq a_{c,l}
\end{cases}
\,.
\end{align}
where
\begin{align}
 a_{g,l} \equiv \frac{G M_{\mathrm{gas},l}}{R_c^2} \,, \quad
 a_{c,l} \equiv \frac{G (M_{\mathrm{gas},l} + M_{\ast,l})}{R_c^2} \,,
\end{align}
where $M_{\ast,l}$ is the stellar mass of the lens $l$, $M_{\mathrm{gas},l}$ is the hot gas mass of the lens $l$, and $R_c = 100\,\mathrm{kpc}$ is where we cut off the SIS profile.

\begin{figure}
 \centering
  \includegraphics[width=.8\textwidth]{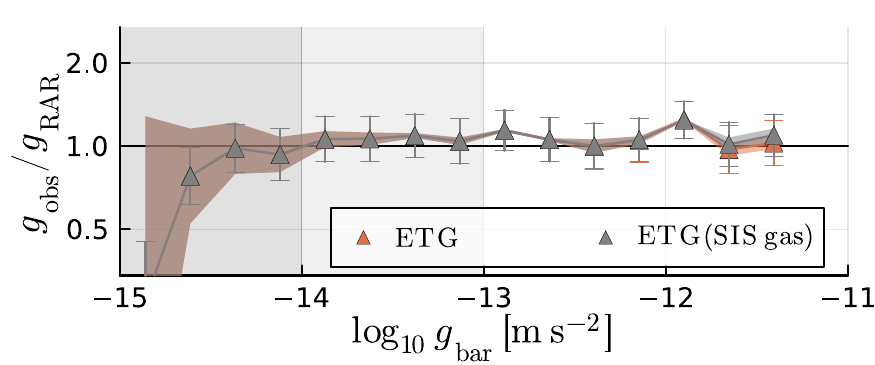}
  \caption{
    Same as Fig.~\ref{fig:etgltg-Risol-4.0}, but we additionally show the result for ETGs with their surrounding hot gas modeled as an SIS cut off at $100\,\mathrm{kpc}$ instead of a point mass (gray color).
    This makes a small difference at relatively large $g_{\mathrm{bar}}$ and does not change the result at small $g_{\mathrm{bar}}$.
    For clarity, we do not show lensing data for LTGs and kinematic data.
  }
 \label{fig:etgltg-hotgas-SIS}
\end{figure}

Fig.~\ref{fig:etgltg-hotgas-SIS} shows that modeling the hot gas of ETGs as an SIS cut off at $100\,\mathrm{kpc}$ instead of a point mass has a small effect at relatively large $g_{\mathrm{bar}}$ and leaves the results unchanged at small $g_{\mathrm{bar}}$.
These results are due the SIS profile being less concentrated towards the center than a point particle.
There would be a larger effect at large $\gbar$ for gas profiles that are even less concentrated towards the center.
For profiles that are more concentrated towards the center, such as an NFW profile, we would expect a smaller effect.
At sufficiently small $\gbar$, where all the gas mass is enclosed so that $\gbar$ can be approximated to fall off as $1/r^2$, the effect will always be negligible, irrespective of the gas profile at larger $\gbar$.

\section{Comparison to Ref.~\cite{Brouwer2021}}

\subsection{SIS method}
\label{sec:appendix:SIScompare}

\begin{figure}
 \centering
  \includegraphics[width=.9\textwidth]{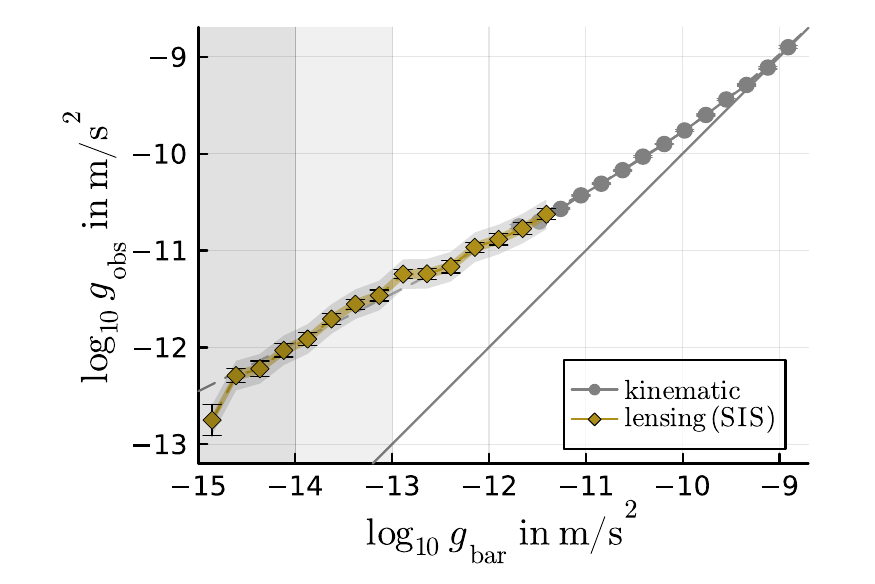}
 \caption{
    Same as Fig.~\ref{fig:rar} but with $g_{\mathrm{obs}}$ derived using the SIS approximation proposed in Ref.~\cite{Brouwer2021} instead of our method based on the exact deprojection formula Eq.~\eqref{eq:gobs_from_esd}.
    The SIS method gives a similar result as our method, but the resulting RAR is less smooth and generally has a larger systematic uncertainty of $0.05\,\mathrm{dex}$ from converting ESD profiles to radial accelerations (see Sec.~\ref{sec:newmethod:individual}).
 }
 \label{fig:rar-SIS-QETG-1.4}
\end{figure}

Here, we compare the weak-lensing RAR derived using the SIS method proposed by Ref.~\cite{Brouwer2021} to that derived using our method based on the exact deprojection formula Eq.~\eqref{eq:gobs_from_esd} (see Sec.~\ref{sec:newmethod}).
Fig.~\ref{fig:rar-SIS-QETG-1.4} shows the RAR derived using our lens sample but with the SIS method (i.e. using Eq.~\eqref{eq:gobs_sis}).
Comparing to Fig.~\ref{fig:rar} we see that both methods produce similar results, but the RAR produced using our new method is smoother and generally comes with a smaller systematic uncertainty.
As discussed in Sec.~\ref{sec:newmethod}, the systematic uncertainty using our new method becomes significant only close to the last data point.

Our lens sample satisfies a stricter isolation criterion than that of Ref.~\cite{Brouwer2021}, namely $R_{\mathrm{isol}} = 4\,\mathrm{Mpc}/h_{70}$ rather than $R_{\mathrm{isol}} = 3\,\mathrm{Mpc}/h_{70}$.
This means a cleaner lens sample, but also a smaller sample and thus larger statistical fluctuations.
Due to the integral in Eq.~\eqref{eq:gobs_from_esd}, our improved method still produces quite smooth results even with such reduced statistics.

\subsection{ETGs vs LTGs}
\label{sec:appendix:BrouwerETGLTG}

\begin{figure}
 \centering
  \includegraphics[width=.8\textwidth]{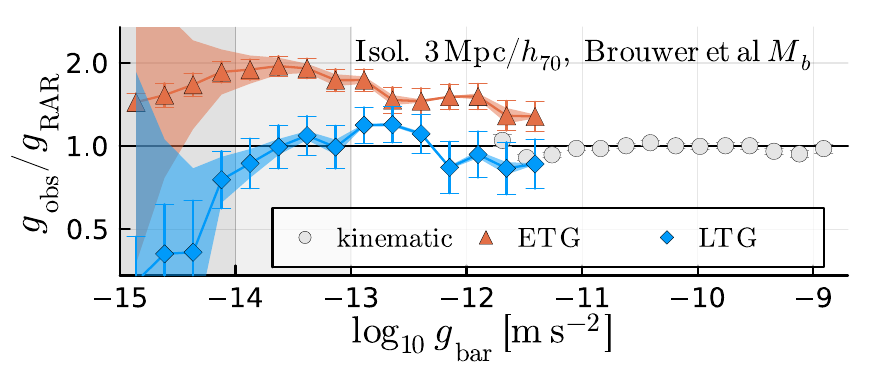}
  \caption{
    Same as Fig.~\ref{fig:etgltg-Risol-4.0}, but \emph{not} using our preferred baryonic masses and isolation criterion.
    Instead, we here adopt $R_{\mathrm{isol}} = 3\,\mathrm{Mpc}/h_{70}$, $Q_{\mathrm{ETG}} = 1.0$, and the gas mass estimate of Ref.~\cite{Brouwer2021} in order to show that we can reproduce the finding of Ref.~\cite{Brouwer2021}.
 }
 \label{fig:etgltg-Risol-3.0-Brouwer-Mb}
\end{figure}

In Fig.~\ref{fig:etgltg-Risol-3.0-Brouwer-Mb}, we demonstrate that we can reproduce the finding of Ref.~\cite{Brouwer2021} that the RAR of ETGs and LTGs differs significantly when we adopt the same baryonic masses and isolation criterion.
Specifically, for a reasonably direct comparison with Ref.~\cite{Brouwer2021}, Fig.~\ref{fig:etgltg-Risol-3.0-Brouwer-Mb} uses $R_{\mathrm{isol}} = 3\,\mathrm{Mpc}/h_{70}$, $Q_{\mathrm{ETG}} = 1.0$, and the gas mass estimate from Ref.~\cite{Brouwer2021} rather than our default choices $R_{\mathrm{isol}} = 4\,\mathrm{Mpc}/h_{70}$, $Q_{\mathrm{ETG}} = 1.4$, and our gas mass estimate from Sec.~\ref{sec:masses}.

\begin{figure}
 \centering
  \includegraphics[width=.8\textwidth]{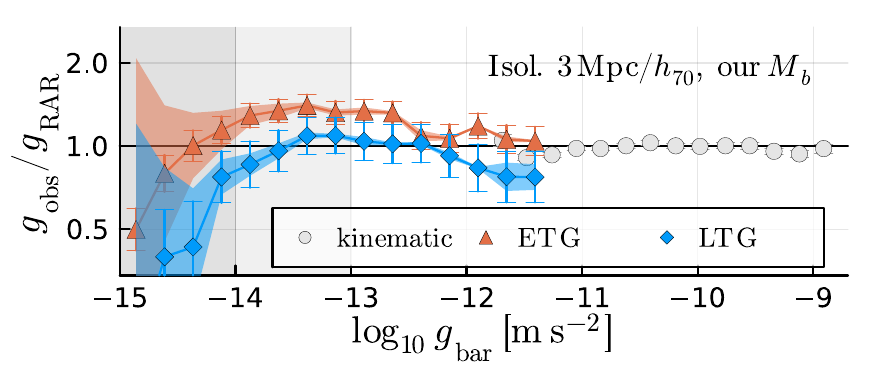}
  \caption{
    As Fig.~\ref{fig:etgltg-Risol-3.0-Brouwer-Mb} but using our baryonic mass estimates from Sec.~\ref{sec:masses} (i.e. using $Q_{\mathrm{ETG}} = 1.4$ and the gas mass estimates Eq.~\eqref{eq:fcold} and Eq.~\eqref{eq:fhot}).
    This brings the RAR of ETGs and LTGs closer together, but does not eliminate the difference.
 }
 \label{fig:etgltg-Risol-3.0}
\end{figure}

In Sec.~\ref{sec:etgltg} we discuss in detail how this result changes when adopting our baryonic masses and our isolation criterion.
In Fig.~\ref{fig:etgltg-Risol-3.0}, we show that adopting our baryonic masses ($Q_{\mathrm{ETG}} = 1.4$ and the gas mass estimates Eq.~\eqref{eq:fcold} and Eq.~\eqref{eq:fhot}) while keeping the isolation criterion of Ref.~\cite{Brouwer2021} ($R_{\mathrm{isol}} = 3\,\mathrm{Mpc}/h_{70}$) reduces the difference between ETGs and LTGs but does not eliminate it.

\bibliographystyle{JHEP}
\bibliography{lensing-RAR.bib}

\end{document}

%% file: plots/RAR-table.tex
$-11.41$ & $-10.65$ & $0.06$ & $0.03$\\
$-11.65$ & $-10.78$ & $0.06$ & $0.03$\\
$-11.90$ & $-10.88$ & $0.06$ & $0.00$\\
$-12.15$ & $-11.00$ & $0.06$ & $0.00$\\
$-12.39$ & $-11.11$ & $0.05$ & $0.02$\\
$-12.64$ & $-11.21$ & $0.05$ & $0.00$\\
$-12.89$ & $-11.29$ & $0.05$ & $0.01$\\
$-13.13$ & $-11.47$ & $0.05$ & $0.02$\\
$-13.38$ & $-11.59$ & $0.05$ & $0.01$\\
$-13.63$ & $-11.76$ & $0.06$ & $0.03$\\
$-13.87$ & $-11.93$ & $0.07$ & $0.05$\\
$-14.12$ & $-12.08$ & $0.07$ & $0.07$\\
$-14.37$ & $-12.27$ & $0.08$ & $0.13$\\
$-14.61$ & $-12.44$ & $0.08$ & $0.25$\\
$-14.86$ & $-12.85$ & $0.12$ & $0.67$\\

%% file: lensing-RAR.bbl
\providecommand{\href}[2]{#2}\begingroup\raggedright\begin{thebibliography}{10}

\bibitem{Brouwer2021}
M.M.~{Brouwer}, K.A.~{Oman}, E.A.~{Valentijn}, M.~{Bilicki}, C.~{Heymans},
  H.~{Hoekstra} et~al., \emph{{The weak lensing radial acceleration relation:
  Constraining modified gravity and cold dark matter theories with KiDS-1000}},
  \href{https://doi.org/10.1051/0004-6361/202040108}{\emph{Astronomy \&
  Astrophysics} {\bfseries 650} A113}
  [\href{https://arxiv.org/abs/2106.11677}{{\ttfamily 2106.11677}}].

\bibitem{Rubin1978}
V.C.~{Rubin}, J.~{Ford}, W.~K. and N.~{Thonnard}, \emph{{Extended rotation
  curves of high-luminosity spiral galaxies. IV. Systematic dynamical
  properties, Sa -> Sc.}}, \href{https://doi.org/10.1086/182804}{\emph{\apjl}
  {\bfseries 225} L107–L111}.

\bibitem{Bosma1981}
A.~{Bosma}, \emph{{21-cm line studies of spiral galaxies. II. The distribution
  and kinematics of neutral hydrogen in spiral galaxies of various
  morphological types.}}, \href{https://doi.org/10.1086/113063}{\emph{\aj}
  {\bfseries 86} 1825–1846}.

\bibitem{McGaugh2000}
S.S.~{McGaugh}, J.M.~{Schombert}, G.D.~{Bothun} and W.J.G.~{ de Blok},
  \emph{{The Baryonic Tully-Fisher Relation}},
  \href{https://doi.org/10.1086/312628}{\emph{Astrophys. J. Lett.} {\bfseries
  533} (2000) L99–L102}
  [\href{https://arxiv.org/abs/astro-ph/0003001}{{\ttfamily
  astro-ph/0003001}}].

\bibitem{McGaugh2016}
S.S.~{McGaugh}, \emph{{The Surface Density Profile of the Galactic Disk from
  the Terminal Velocity Curve}},
  \href{https://doi.org/10.3847/0004-637X/816/1/42}{\emph{Astrophys. J.}
  {\bfseries 816} 42} [\href{https://arxiv.org/abs/1511.06387}{{\ttfamily
  1511.06387}}].

\bibitem{Lelli2016c}
F.~{Lelli}, S.S.~{McGaugh} and J.M.~{Schombert}, \emph{{The Small Scatter of
  the Baryonic Tully-Fisher Relation}},
  \href{https://doi.org/10.3847/2041-8205/816/1/L14}{\emph{\apjl} {\bfseries
  816} L14} [\href{https://arxiv.org/abs/1512.04543}{{\ttfamily 1512.04543}}].

\bibitem{Lelli2019}
F.~{Lelli}, S.S.~{McGaugh}, J.M.~{Schombert}, H.~{Desmond} and H.~{Katz},
  \emph{{The baryonic Tully-Fisher relation for different velocity definitions
  and implications for galaxy angular momentum}},
  \href{https://doi.org/10.1093/mnras/stz205}{\emph{\mnras} {\bfseries 484}
  3267–3278} [\href{https://arxiv.org/abs/1901.05966}{{\ttfamily
  1901.05966}}].

\bibitem{Schombert2020}
J.~{Schombert}, S.~{McGaugh} and F.~{Lelli}, \emph{{Using the Baryonic
  Tully-Fisher Relation to Measure H$_{o}$}},
  \href{https://doi.org/10.3847/1538-3881/ab9d88}{\emph{Astronom. J.}
  {\bfseries 160} 71} [\href{https://arxiv.org/abs/2006.08615}{{\ttfamily
  2006.08615}}].

\bibitem{Lelli2016b}
F.~{Lelli}, S.S.~{McGaugh}, J.M.~{Schombert} and M.S.~{Pawlowski}, \emph{{The
  Relation between Stellar and Dynamical Surface Densities in the Central
  Regions of Disk Galaxies}},
  \href{https://doi.org/10.3847/2041-8205/827/1/L19}{\emph{Astrophys. J. Lett.}
  {\bfseries 827} (2016) L19}
  [\href{https://arxiv.org/abs/1607.02145}{{\ttfamily 1607.02145}}].

\bibitem{Milgrom2016b}
M.~{Milgrom}, \emph{{Universal Modified Newtonian Dynamics Relation between the
  Baryonic and “Dynamical” Central Surface Densities of Disc Galaxies}},
  \href{https://doi.org/10.1103/PhysRevLett.117.141101}{\emph{Phys. Rev. Lett.}
  {\bfseries 117} (2016) 141101}
  [\href{https://arxiv.org/abs/1607.05103}{{\ttfamily 1607.05103}}].

\bibitem{Sanders1990}
R.H.~{Sanders}, \emph{{Mass discrepancies in galaxies: dark matter and
  alternatives}}, \href{https://doi.org/10.1007/BF00873540}{\emph{\aapr}
  {\bfseries 2} 1–28}.

\bibitem{McGaugh2004b}
S.S.~{McGaugh}, \emph{{The Mass Discrepancy-Acceleration Relation: Disk Mass
  and the Dark Matter Distribution}},
  \href{https://doi.org/10.1086/421338}{\emph{\apj} {\bfseries 609} 652–666}
  [\href{https://arxiv.org/abs/astro-ph/0403610}{{\ttfamily
  astro-ph/0403610}}].

\bibitem{Wu2015}
X.~{Wu} and P.~{Kroupa}, \emph{{Galactic rotation curves, the
  baryon-to-dark-halo-mass relation and space-time scale invariance}},
  \href{https://doi.org/10.1093/mnras/stu2099}{\emph{\mnras} {\bfseries 446}
  330–344} [\href{https://arxiv.org/abs/1410.2256}{{\ttfamily 1410.2256}}].

\bibitem{McGaugh2016b}
S.S.~{McGaugh}, F.~{Lelli} and J.M.~{Schombert}, \emph{{Radial Acceleration
  Relation in Rotationally Supported Galaxies}},
  \href{https://doi.org/10.1103/PhysRevLett.117.201101}{\emph{\prl} {\bfseries
  117} 201101} [\href{https://arxiv.org/abs/1609.05917}{{\ttfamily
  1609.05917}}].

\bibitem{Lelli2017b}
F.~{Lelli}, S.S.~{McGaugh}, J.M.~{Schombert} and M.S.~{Pawlowski}, \emph{{One
  Law to Rule Them All: The Radial Acceleration Relation of Galaxies}},
  \href{https://doi.org/10.3847/1538-4357/836/2/152}{\emph{Astrophys. J.}
  {\bfseries 836} 152} [\href{https://arxiv.org/abs/1610.08981}{{\ttfamily
  1610.08981}}].

\bibitem{Shelest2020}
A.~{Shelest} and F.~{Lelli}, \emph{{From spirals to lenticulars: Evidence from
  the rotation curves and mass models of three early-type galaxies}},
  \href{https://doi.org/10.1051/0004-6361/202038184}{\emph{\aap} {\bfseries
  641} A31} [\href{https://arxiv.org/abs/2006.10813}{{\ttfamily 2006.10813}}].

\bibitem{Tully1977}
R.B.~{Tully} and J.R.~{Fisher}, \emph{{A new method of determining distances to
  galaxies.}}, {\emph{\aap} {\bfseries 54} 661–673}.

\bibitem{Verheijen2001}
M.A.W.~{Verheijen}, \emph{{The Ursa Major Cluster of Galaxies. V. H I Rotation
  Curve Shapes and the Tully-Fisher Relations}},
  \href{https://doi.org/10.1086/323887}{\emph{\apj} {\bfseries 563} (2001) 694}
  [\href{https://arxiv.org/abs/astro-ph/0108225}{{\ttfamily
  astro-ph/0108225}}].

\bibitem{Ponomareva2017}
A.A.~{Ponomareva}, M.A.W.~{Verheijen}, R.F.~{ Peletier} and A.~{Bosma},
  \emph{{The multiwavelength Tully-Fisher relation with spatially resolved H I
  kinematics}}, \href{https://doi.org/10.1093/mnras/stx1018}{\emph{\mnras}
  {\bfseries 469} (2017) 2387}
  [\href{https://arxiv.org/abs/1704.08788}{{\ttfamily 1704.08788}}].

\bibitem{Ponomareva2018}
A.A.~{Ponomareva}, M.A.W.~{Verheijen}, E.~{ Papastergis}, A.~{Bosma} and
  R.F.~{Peletier}, \emph{{From light to baryonic mass: the effect of the
  stellar mass-to-light ratio on the Baryonic Tully-Fisher relation}},
  \href{https://doi.org/10.1093/mnras/stx3066}{\emph{\mnras} {\bfseries 474}
  (2018) 4366} [\href{https://arxiv.org/abs/1711.09112}{{\ttfamily
  1711.09112}}].

\bibitem{Lelli2016}
F.~{Lelli}, S.S.~{McGaugh} and J.M.~{Schombert}, \emph{{SPARC: Mass Models for
  175 Disk Galaxies with Spitzer Photometry and Accurate Rotation Curves}},
  \href{https://doi.org/10.3847/0004-6256/152/6/157}{\emph{The Astronomical
  Journal} {\bfseries 152} 157}
  [\href{https://arxiv.org/abs/1606.09251}{{\ttfamily 1606.09251}}].

\bibitem{Noordermeer2005}
E.~{Noordermeer}, J.M.~{van der Hulst}, R.~{Sancisi}, R.A.~{Swaters} and
  T.S.~{van Albada}, \emph{{The Westerbork HI survey of spiral and irregular
  galaxies. III. HI observations of early-type disk galaxies}},
  \href{https://doi.org/10.1051/0004-6361:20053172}{\emph{\aap} {\bfseries 442}
  137–157} [\href{https://arxiv.org/abs/astro-ph/0508319}{{\ttfamily
  astro-ph/0508319}}].

\bibitem{Lelli2010}
F.~{Lelli}, F.~{Fraternali} and R.~{Sancisi}, \emph{{Structure and dynamics of
  giant low surface brightness galaxies}},
  \href{https://doi.org/10.1051/0004-6361/200913808}{\emph{\aap} {\bfseries
  516} A11} [\href{https://arxiv.org/abs/1003.1312}{{\ttfamily 1003.1312}}].

\bibitem{Brimioulle2013}
F.~{Brimioulle}, S.~{Seitz}, M.~{Lerchster}, R.~{Bender} and J.~{Snigula},
  \emph{{Dark matter halo properties from galaxy-galaxy lensing}},
  \href{https://doi.org/10.1093/mnras/stt525}{\emph{\mnras} {\bfseries 432}
  (2013) 1046–1102} [\href{https://arxiv.org/abs/1303.6287}{{\ttfamily
  1303.6287}}].

\bibitem{Milgrom2013}
M.~{Milgrom}, \emph{{Testing the MOND Paradigm of Modified Dynamics with
  Galaxy-Galaxy Gravitational Lensing}},
  \href{https://doi.org/10.1103/PhysRevLett.111.041105}{\emph{\prl} {\bfseries
  111} 041105} [\href{https://arxiv.org/abs/1305.3516}{{\ttfamily 1305.3516}}].

\bibitem{Brouwer2017}
M.M.~{Brouwer}, M.R.~{Visser}, A.~{Dvornik}, H.~{Hoekstra}, K.~{Kuijken},
  E.A.~{Valentijn} et~al., \emph{{First test of Verlinde's theory of emergent
  gravity using weak gravitational lensing measurements}},
  \href{https://doi.org/10.1093/mnras/stw3192}{\emph{\mnras} {\bfseries 466}
  (2017) 2547–2559} [\href{https://arxiv.org/abs/1612.03034}{{\ttfamily
  1612.03034}}].

\bibitem{Kuijken2019}
K.~{Kuijken}, C.~{Heymans}, A.~{Dvornik}, H.~{Hildebrandt }, J.T.A.~{de Jong},
  A.H.~{Wright} et~al., \emph{{The fourth data release of the Kilo-Degree
  Survey: ugri imaging and nine-band optical-IR photometry over 1000 square
  degrees}}, \href{https://doi.org/10.1051/0004-6361/201834918}{\emph{\aap}
  {\bfseries 625} (2019) A2}
  [\href{https://arxiv.org/abs/1902.11265}{{\ttfamily 1902.11265}}].

\bibitem{Giblin2021}
B.~{Giblin}, C.~{Heymans}, M.~{Asgari}, H.~{Hildebrandt}, H.~{Hoekstra},
  B.~{Joachimi} et~al., \emph{{KiDS-1000 catalogue: Weak gravitational lensing
  shear measurements }},
  \href{https://doi.org/10.1051/0004-6361/202038850}{\emph{\aap} {\bfseries
  645} (2021) A105} [\href{https://arxiv.org/abs/2007.01845}{{\ttfamily
  2007.01845}}].

\bibitem{Bilicki2021}
M.~{Bilicki}, A.~{Dvornik}, H.~{Hoekstra}, A.H.~{Wright}, N.E.~{Chisari},
  M.~{Vakili} et~al., \emph{{Bright galaxy sample in the Kilo-Degree Survey
  Data Release 4. Selection, photometric redshifts, and physical properties}},
  \href{https://doi.org/10.1051/0004-6361/202140352}{\emph{\aap} {\bfseries
  653} (2021) A82}.

\bibitem{Schombert2014}
J.~{Schombert} and S.~{McGaugh}, \emph{{Stellar Populations and the Star
  Formation Histories of LSB Galaxies: III. Stellar Population Models}},
  \href{https://doi.org/10.1017/pasa.2014.32}{\emph{\pasa} {\bfseries 31} e036}
  [\href{https://arxiv.org/abs/1407.6778}{{\ttfamily 1407.6778}}].

\bibitem{Kaiser1995}
N.~{Kaiser}, \emph{{Nonlinear Cluster Lens Reconstruction}},
  \href{https://doi.org/10.1086/187730}{\emph{\apjl} {\bfseries 439} (1995) L1}
  [\href{https://arxiv.org/abs/astro-ph/9408092}{{\ttfamily
  astro-ph/9408092}}].

\bibitem{Bartelmann1995}
M.~{Bartelmann}, \emph{{Cluster mass estimates from weak lensing.}},
  \href{https://doi.org/10.48550/arXiv.astro-ph/9412051}{\emph{\aap} {\bfseries
  303} (1995) 643} [\href{https://arxiv.org/abs/astro-ph/9412051}{{\ttfamily
  astro-ph/9412051}}].

\bibitem{Bartelmann2001}
M.~{Bartelmann} and P.~{Schneider}, \emph{{Weak gravitational lensing}},
  \href{https://doi.org/10.1016/S0370-1573(00)00082-X}{\emph{\physrep}
  {\bfseries 340} 291–472}
  [\href{https://arxiv.org/abs/astro-ph/9912508}{{\ttfamily
  astro-ph/9912508}}].

\bibitem{Mistele2023}
T.~{Mistele}, S.~{McGaugh} and S.~{Hossenfelder}, \emph{{Aether scalar tensor
  theory confronted with weak lensing data at small accelerations}},
  \href{https://doi.org/10.1051/0004-6361/202346025}{\emph{\aap} {\bfseries
  676} (2023) A100} [\href{https://arxiv.org/abs/2301.03499}{{\ttfamily
  2301.03499}}].

\bibitem{Mistele2023c}
T.~{Mistele}, S.~{McGaugh} and S.~{Hossenfelder}, \emph{{Superfluid dark matter
  in tension with weak gravitational lensing data}},
  \href{https://doi.org/10.1088/1475-7516/2023/09/004}{\emph{\jcap} {\bfseries
  2023} (2023) 004} [\href{https://arxiv.org/abs/2303.08560}{{\ttfamily
  2303.08560}}].

\bibitem{Skordis2020}
C.~{Skordis} and T.~{Z{ł}osnik}, \emph{{New Relativistic Theory for Modified
  Newtonian Dynamics}},
  \href{https://doi.org/10.1103/PhysRevLett.127.161302}{\emph{Phys. Rev. Lett.}
  {\bfseries 127} 161302} [\href{https://arxiv.org/abs/2007.00082}{{\ttfamily
  2007.00082}}].

\bibitem{Milgrom1986b}
M.~{Milgrom}, \emph{{Can the Hidden Mass Be Negative?}},
  \href{https://doi.org/10.1086/164314}{\emph{\apj} {\bfseries 306} 9}.

\bibitem{Wright2020}
A.H.~{Wright}, H.~{Hildebrandt}, J.L.~{van den Busch} and C.~{Heymans},
  \emph{{Photometric redshift calibration with self-organising maps}},
  \href{https://doi.org/10.1051/0004-6361/201936782}{\emph{\aap} {\bfseries
  637} (2020) A100} [\href{https://arxiv.org/abs/1909.09632}{{\ttfamily
  1909.09632}}].

\bibitem{Hildebrandt2021}
H.~{Hildebrandt}, J.L.~{van den Busch}, A.H.~{Wright}, C.~{Blake},
  B.~{Joachimi}, K.~{Kuijken} et~al., \emph{{KiDS-1000 catalogue: Redshift
  distributions and their calibration}},
  \href{https://doi.org/10.1051/0004-6361/202039018}{\emph{\aap} {\bfseries
  647} (2021) A124} [\href{https://arxiv.org/abs/2007.15635}{{\ttfamily
  2007.15635}}].

\bibitem{Sadeh2016}
I.~{Sadeh}, F.B.~{Abdalla} and O.~{Lahav}, \emph{{ANNz2: Photometric Redshift
  and Probability Distribution Function Estimation using Machine Learning}},
  \href{https://doi.org/10.1088/1538-3873/128/968/104502}{\emph{\pasp}
  {\bfseries 128} (2016) 104502}
  [\href{https://arxiv.org/abs/1507.00490}{{\ttfamily 1507.00490}}].

\bibitem{Schombert2019}
J.~{Schombert}, S.~{McGaugh} and F.~{Lelli}, \emph{{The mass-to-light ratios
  and the star formation histories of disc galaxies}},
  \href{https://doi.org/10.1093/mnras/sty3223}{\emph{\mnras} {\bfseries 483}
  (2019) 1496} [\href{https://arxiv.org/abs/1811.10579}{{\ttfamily
  1811.10579}}].

\bibitem{Dvornik2017}
A.~{Dvornik}, M.~{Cacciato}, K.~{Kuijken}, M.~{Viola}, H.~{Hoekstra},
  R.~{Nakajima} et~al., \emph{{A KiDS weak lensing analysis of assembly bias in
  GAMA galaxy groups}},
  \href{https://doi.org/10.1093/mnras/stx705}{\emph{\mnras} {\bfseries 468}
  (2017) 3251–3265} [\href{https://arxiv.org/abs/1703.06657}{{\ttfamily
  1703.06657}}].

\bibitem{McGaugh2014}
S.S.~{McGaugh} and J.M.~{Schombert}, \emph{{Color-Mass-to-light-ratio Relations
  for Disk Galaxies}},
  \href{https://doi.org/10.1088/0004-6256/148/5/77}{\emph{\aj} {\bfseries 148}
  (2014) 77} [\href{https://arxiv.org/abs/1407.1839}{{\ttfamily 1407.1839}}].

\bibitem{Pengfei2019}
P.~{Li}, F.~{Lelli}, S.~{McGaugh}, M.S.~{ Pawlowski}, M.A.~{Zwaan} and
  J.~{Schombert}, \emph{{The Halo Mass Function of Late-type Galaxies from H I
  Kinematics}}, \href{https://doi.org/10.3847/2041-8213/ab53e6}{\emph{\apjl}
  {\bfseries 886} (2019) L11}
  [\href{https://arxiv.org/abs/1911.00517}{{\ttfamily 1911.00517}}].

\bibitem{CG2010}
C.~{Conroy} and J.E.~{Gunn}, \emph{{The Propagation of Uncertainties in Stellar
  Population Synthesis Modeling. III. Model Calibration, Comparison, and
  Evaluation}}, \href{https://doi.org/10.1088/0004-637X/712/2/833}{\emph{\apj}
  {\bfseries 712} (2010) 833}
  [\href{https://arxiv.org/abs/0911.3151}{{\ttfamily 0911.3151}}].

\bibitem{BC2003}
G.~{Bruzual} and S.~{Charlot}, \emph{{Stellar population synthesis at the
  resolution of 2003}},
  \href{https://doi.org/10.1046/j.1365-8711.2003.06897.x}{\emph{\mnras}
  {\bfseries 344} (2003) 1000}
  [\href{https://arxiv.org/abs/astro-ph/0309134}{{\ttfamily
  astro-ph/0309134}}].

\bibitem{Schombert2022}
J.~{Schombert}, S.~{McGaugh} and F.~{Lelli}, \emph{{Stellar Mass-to-light
  Ratios: Composite Bulge+Disk Models and the Baryonic Tully-Fisher Relation}},
  \href{https://doi.org/10.3847/1538-3881/ac5249}{\emph{\aj} {\bfseries 163}
  (2022) 154} [\href{https://arxiv.org/abs/2202.02290}{{\ttfamily
  2202.02290}}].

\bibitem{Kodama1997}
T.~{Kodama} and N.~{Arimoto}, \emph{{Origin of the colour-magnitude relation of
  elliptical galaxies.}},
  \href{https://doi.org/10.48550/arXiv.astro-ph/9609160}{\emph{\aap} {\bfseries
  320} (1997) 41} [\href{https://arxiv.org/abs/astro-ph/9609160}{{\ttfamily
  astro-ph/9609160}}].

\bibitem{Buote2012}
D.A.~{Buote} and P.J.~{Humphrey}, \emph{{Dark Matter in Elliptical Galaxies}},
  in \emph{{Astrophysics and Space Science Library}}, D.-W.~{Kim} and
  S.~{Pellegrini}, eds., vol.~378 of \emph{{Astrophysics and Space Science
  Library}}, p.~235, Jan., 2012,
  \href{https://doi.org/10.1007/978-1-4614-0580-1_8}{DOI}
  [\href{https://arxiv.org/abs/1104.0012}{{\ttfamily 1104.0012}}].

\bibitem{Chae2021b}
K.-H.~{Chae}, H.~{Desmond}, F.~{Lelli}, S.S.~{ McGaugh} and J.M.~{Schombert},
  \emph{{Testing the Strong Equivalence Principle. II. Relating the External
  Field Effect in Galaxy Rotation Curves to the Large-scale Structure of the
  Universe}}, \href{https://doi.org/10.3847/1538-4357/ac1bba}{\emph{Astrophys.
  J.} {\bfseries 921} (2021) 104}
  [\href{https://arxiv.org/abs/2109.04745}{{\ttfamily 2109.04745}}].

\bibitem{McGaugh2020b}
S.S.~{McGaugh}, F.~{Lelli} and J.M.~{Schombert}, \emph{{Scaling Relations for
  Molecular Gas and Metallicity: Impact on the Baryonic Tully-Fisher
  Relation}}, \href{https://doi.org/10.3847/2515-5172/ab8471}{\emph{Research
  Notes of the American Astronomical Society} {\bfseries 4} 45}.

\bibitem{McGaugh2011}
S.S.~{McGaugh}, \emph{{Novel Test of Modified Newtonian Dynamics with Gas Rich
  Galaxies}}, \href{https://doi.org/10.1103/PhysRevLett.106.121303}{\emph{\prl}
  {\bfseries 106} 121303} [\href{https://arxiv.org/abs/1102.3913}{{\ttfamily
  1102.3913}}].

\bibitem{Dressler1980}
A.~{Dressler}, \emph{{Galaxy morphology in rich clusters: implications for the
  formation and evolution of galaxies.}},
  \href{https://doi.org/10.1086/157753}{\emph{\apj} {\bfseries 236} 351–365}.

\bibitem{Cappellari2012}
M.~{Cappellari}, R.M.~{McDermid}, K.~{Alatalo}, L.~{Blitz}, M.~{Bois},
  F.~{Bournaud} et~al., \emph{{Systematic variation of the stellar initial mass
  function in early-type galaxies}},
  \href{https://doi.org/10.1038/nature10972}{\emph{\nat} {\bfseries 484} (2012)
  485} [\href{https://arxiv.org/abs/1202.3308}{{\ttfamily 1202.3308}}].

\bibitem{Conroy2012}
C.~{Conroy} and P.G.~{van Dokkum}, \emph{{The Stellar Initial Mass Function in
  Early-type Galaxies From Absorption Line Spectroscopy. II. Results}},
  \href{https://doi.org/10.1088/0004-637X/760/1/71}{\emph{\apj} {\bfseries 760}
  (2012) 71} [\href{https://arxiv.org/abs/1205.6473}{{\ttfamily 1205.6473}}].

\bibitem{Conroy2013}
C.~{Conroy}, A.A.~{Dutton}, G.J.~{Graves}, J.T.~{Mendel} and P.G.~{van Dokkum},
  \emph{{Dynamical versus Stellar Masses in Compact Early-type Galaxies:
  Further Evidence for Systematic Variation in the Stellar Initial Mass
  Function}}, \href{https://doi.org/10.1088/2041-8205/776/2/L26}{\emph{\apjl}
  {\bfseries 776} (2013) L26}
  [\href{https://arxiv.org/abs/1306.2316}{{\ttfamily 1306.2316}}].

\bibitem{Li2022}
P.~{Li}, S.S.~{McGaugh}, F.~{Lelli}, Y.~{ Tian}, J.M.~{Schombert} and
  C.-M.~{Ko}, \emph{{The Effect of Adiabatic Compression on Dark Matter Halos
  and the Radial Acceleration Relation}},
  \href{https://doi.org/10.3847/1538-4357/ac52aa}{\emph{\apj} {\bfseries 927}
  (2022) 198} [\href{https://arxiv.org/abs/2202.03421}{{\ttfamily
  2202.03421}}].

\bibitem{Mercado2023}
F.J.~{Mercado}, J.S.~{Bullock}, J.~{Moreno}, M.~{Boylan-Kolchin}, A.~{Wetzel},
  C.-A.~{ Faucher-Giguère} et~al., \emph{{Hooks \& Bends in the Radial
  Acceleration Relation: Tests for Dark Matter and Challenges for MOND}},
  {\emph{arXiv e-prints} (2023) }
  [\href{https://arxiv.org/abs/2307.09507}{{\ttfamily 2307.09507}}].

\bibitem{Milgrom1983a}
M.~{Milgrom}, \emph{{A Modification of the Newtonian dynamics as a possible
  alternative to the hidden mass hypothesis}},
  \href{https://doi.org/10.1086/161130}{\emph{Astrophys. J.} {\bfseries 270}
  (1983) 365–370}.

\bibitem{Milgrom1983b}
M.~{Milgrom}, \emph{{A modification of the Newtonian dynamics - Implications
  for galaxies.}}, \href{https://doi.org/10.1086/161131}{\emph{Astrophys. J.}
  {\bfseries 270} (1983) 371–383}.

\bibitem{Milgrom1983c}
M.~{Milgrom}, \emph{{A modification of the Newtonian dynamics: implications for
  galaxy systems}}, \href{https://doi.org/10.1086/161132}{\emph{Astrophys. J.}
  {\bfseries 270} (1983) 384–389}.

\bibitem{Famaey2012}
B.~{Famaey} and S.S.~{McGaugh}, \emph{{Modified Newtonian Dynamics (MOND):
  Observational Phenomenology and Relativistic Extensions}},
  \href{https://doi.org/10.12942/lrr-2012-10}{\emph{Living Reviews in
  Relativity} {\bfseries 15} (2012) 10}
  [\href{https://arxiv.org/abs/1112.3960}{{\ttfamily 1112.3960}}].

\bibitem{Chae2022b}
K.-H.~{Chae} and M.~{Milgrom}, \emph{{Numerical Solutions of the External Field
  Effect on the Radial Acceleration in Disk Galaxies}},
  \href{https://doi.org/10.3847/1538-4357/ac5405}{\emph{\apj} {\bfseries 928}
  (2022) 24} [\href{https://arxiv.org/abs/2201.02109}{{\ttfamily 2201.02109}}].

\bibitem{Kent1986}
S.M.~{Kent}, \emph{{Dark matter in spiral galaxies. I. Galaxies with optical
  rotation curves.}}, \href{https://doi.org/10.1086/114106}{\emph{\aj}
  {\bfseries 91} 1301–1327}.

\bibitem{Mathematica13}
I.~Wolfram~Research{,}, \emph{{Mathematica, {V}ersion 13.3}},  2023.

\bibitem{Viola2015}
M.~{Viola}, M.~{Cacciato}, M.~{Brouwer}, K.~{Kuijken}, H.~{Hoekstra},
  P.~{Norberg} et~al., \emph{{Dark matter halo properties of GAMA galaxy groups
  from 100 square degrees of KiDS weak lensing data}},
  \href{https://doi.org/10.1093/mnras/stv1447}{\emph{\mnras} {\bfseries 452}
  (2015) 3529–3550} [\href{https://arxiv.org/abs/1507.00735}{{\ttfamily
  1507.00735}}].

\end{thebibliography}\endgroup
